 \documentclass[final,3p,times]{elsarticle}

 \usepackage{graphicx}

\usepackage{amssymb}
\usepackage{amsmath}
\usepackage{subfig}

\usepackage{booktabs}

\usepackage{color}

\newcommand{\be}{\begin{equation}}
\newcommand{\ee}{\end{equation}}
\newcommand{\beeq}{\begin{eqnarray}} 
\newcommand{\eeeq}{\end{eqnarray}} 

\newcommand{\uvec}[1]{\underline{#1}}

\begin{document}

\begin{frontmatter}



\title{Gluon cascades and amplitudes in  light-front perturbation theory}


\author{C.A. Cruz-Santiago}
\address{Physics Department, 104 Davey Lab, The Pennsylvania State University, University Park, PA 16802, United States}
\author{A.M. Sta\'sto}
\address{Physics Department, 104 Davey Lab, The Pennsylvania State University, University Park, PA 16802, United States}
\address{RIKEN BNL Center, Upton, New York, United States}
\address{H. Niewodnicza\'nski Institute of Nuclear Physics, Polish Academy of Sciences, ul. Radzikowskiego 152, 31-342 Krak\'ow, Poland}

\begin{abstract}
We construct the gluon wave functions, fragmentation functions and scattering amplitudes within the light-front perturbation theory. Recursion relations on the light-front are constructed for the wave functions and fragmentation functions, which in the latter case are the light-front analogs of the Berends-Giele recursion relations. Using general relations between wave functions
and scattering amplitudes it is demonstrated how to obtain the maximally-helicity violating amplitudes,
and explicit verification of the  results is based on  simple examples.
\end{abstract}

\end{frontmatter}



\section{Introduction}
\label{sec:intro}

Over the past years there has been an enormous progress in the computation of the multi-particle
helicity amplitudes in QCD. The tree-level gluon amplitudes turned out to have relatively simple form
for  special configurations of helicity of external on-shell gluons \cite{Parke:1986gb} despite being the sum of a large number of Feynman diagrams. The recursion relations have been found \cite{Berends:1987me} which allow the computation of  amplitudes with arbitrary number 
of external legs. Presently there exists a number of automated tools for the evaluation of the multi-parton amplitudes  for different species of particles \cite{Mangano:2002ea,Gleisberg:2008fv,Kleiss:2010hy,Cafarella:2007pc,Dixon:2010ik,Alwall:2011uj}.

A natural question thus arises as to whether one can gain more physical insight into the structure of these results.  Quantization on the light-front  \cite{Dirac:1949cp} offers particular advantages which are related to the presence of only three dynamical symmetry operators. In particular, in the infinite momentum frame there exists a remarkable isomorphism between a subgroup of the Poincar\'e group and the Galilean 
symmetry group of non-relativistic quantum mechanics in two dimensions \cite{Susskind:1967rg,Kogut:1972di}.
The vacuum on the light-front is essentially structureless \cite{Weinberg:1966jm} (up to  zero modes), which allows to define unambigously  the partonic content of hadrons and of hadronic wave functions. 
The simplicity of the vacuum has been used to argue about the presence of  in-hadron quark condensates \cite{Brodsky:2010xf}. The light-front framework has been used   to investigate the hadron dynamics  from AdS/CFT correspondence \cite{Brodsky:2003px}. 
Methods of the light-front perturbation theory (LFPT) were used to compute the  soft gluon component of the heavy onium wave function and to 
obtain a  correspondence with the hard Pomeron in QCD \cite{Mueller:1993rr}.
Similarly,  the LFPT has been utilized in the small $x$ limit to analyze the quark distribution function of a heavy meson utilizing ladder operators \cite{Antonuccio:1997tw}. 
In work \cite{ms}  light-front perturbation techniques \cite{Kogut:1969xa,Bjorken:1970ah,Lepage:1980fj,Brodsky:1997de} have been applied to construct the light-front wave function of the gluon.   Expressions for the  recursion relations which are obeyed by these wave functions have been found. They have been solved for some special cases of the helicity configurations. To be precise, an exact form of the tree-level multi gluon components of the real gluon wave function has been found for special helicity configuration in which both the initial gluon and the final gluons carry the same helicities. It turns out that the  variables which occur naturally in the light-front formulation are closely related to the spinor products which are used to express the results for the maximally-helicity-violating (MHV) amplitudes \cite{Parke:1986gb,Mangano:1990by}.
The scattering amplitude for $2\rightarrow n$ process with massless gluons has been also considered in the LFPT, under the special assumption that the evolved projectile gluon is separated in rapidity from the target gluon \cite{ms}. The result in this approximation  has been shown to coincide with the expression of the Parke -Taylor amplitude. 
As an additional result the closed form for the fragmentation function of an off-shell gluon into arbitrary number of final state gluons have been computed. A recursion relation for the fragmentation function has also been found. The recursion involves factorization of the fragmentation trees which is not apparent in the LCPT due to the entanglement of the particle momenta in the energy denominators for the intermediate states. It has been shown that the factorization of the fragmentation trees is 
achieved after the summation of all the light-front time-orderings of all the splittings in the cascade.

The main goal of this paper is to rederive, using the LFPT and techniques presented in \cite{ms} for fragmentation functions and wave functions, the exact expressions for the helicity multi-gluon amplitudes in QCD.
In particular it is important to relax the simplifying assumption used in Ref.~\cite{ms} about the rapidity difference between the projectile and the target gluon.   We first argue  that the factorization property of the fragmentation functions is the light-front analog of the Berends-Giele recursion relations for the gluonic currents. A generalized factorization for the fragmentation function is constructed which includes more complicated helicity configurations.
Next, the explicit relation between the light  front wave functions and scattering amplitudes is shown and verified by the computation of the lowest order examples. Additionally it is shown how the recursion relations for the wave functions in the case of the same helicities for the outgoing gluons are related to the vanishing of the scattering amplitudes with $-,+,\dots,+$ configurations. The vanishing of these amplitudes is related to the kinematical constraint of the conservation of the light-front energy between the initial and final state. The vanishing of the amplitudes with the same helicities for all the gluons is related to the explicit conservation of the total angular momentum on the light-front.  Finally, the recursion relation for the wave functions in the case when one of the outgoing gluons has different helicity is constructed which leads to the recursion relation for the first nontrivial set of MHV amplitudes. Through the explicit computation the Parke-Taylor amplitude is reproduced. 

The structure of the paper is the following: in the next section we briefly summarize the results and techniques derived in the paper \cite{ms}. It is then argued that the factorization for the fragmentation functions is the light-front analog of the Berends-Giele recursion formula. In Sec.~3 the general relations between the light-front wave function and the scattering amplitudes are discussed and examples which confirm this identification are shown. The vanishing of the $+,\dots,+$ and $-,+,\dots,+$ amplitudes on the light-front is demonstrated. The recursion relation for wave function is presented and it is shown that it can be used to rederive the Parke-Taylor amplitude. Finally, in the last section we state conclusions and the outlook for future work.


\section{Gluon wave function and fragmentation function with exact kinematics}
\label{sec:wvfr}

In this section we shall first set up the notation which will be used through the rest of the paper. Next, we will summarize the most important results which were derived in paper \cite{ms}. In particular, 
 we shall show the exact form of the $n$-gluon components of the  gluon wave function, $\Psi_n, \;\;n=1,2,\ldots$ and fragmentation function $T_n$ using the LCPT for a special choice of helicities.  
In the following by the gluon wave function we mean a sub-amplitude on the light-front where the incoming gluon is taken on-shell and the outgoing gluons are all off-shell. Similarly, by the fragmentation function we mean a sub-amplitude with the initial gluon off-shell and the outgoing gluons on-shell. Both wave functions and fragmentation functions can be used as building blocks
to construct the scattering amplitudes and they obey separate recursion relations.

 For now, we shall focus first on the gluon wave functions and as in Ref.~\cite{ms} we  assume that the incoming on-shell gluon has a positive helicity and  all the final off-shell gluons in the wave function have all positive helicities.  The resulting resummed gluon wave function will have a very simple form.

 \begin{figure}[ht]
\centerline{\includegraphics[width=0.4\textwidth]{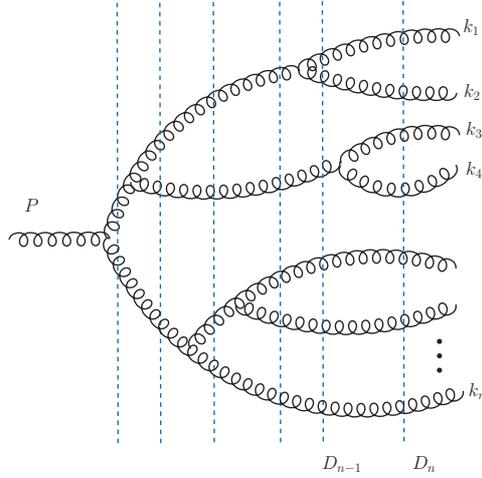}}
\caption{The multi-gluon wave function of the incoming gluon with momentum $P$. The vertical dashed lines symbolize different intermediate states where we need to evaluate the energy denominators. It is understood that the wave function scatters finally on some target. }
\label{fig:onium_n}
\end{figure}
We consider an incoming  gluon
 with the four momentum $P$, and the color index $a$, which evolves into a state containing $n$ gluons with momenta $(k_1,\ldots k_n)$ and color indices $(a_1,a_2,\ldots,a_n)$ correspondingly. This is illustrated in 
 Fig.~\ref{fig:onium_n}. The momentum for the initial virtual gluon with virtuality $-Q^2$ and a  transverse momentum $\uvec{q}$ can be written in the light-front variables, 
\[
P^{\mu} = (z_0 P^+,  \frac{-Q^2+\uvec{q}^2}{ z_0 P^+},\, \uvec{q})\; ,
\]
with $P^{\pm}=P^0\pm P^3$ and $z_0=1$.
The   momenta of the last $n$ - gluons are labeled $k_1,\dots,k_n$, as in  Fig~\ref{fig:onium_n}. Each of this momenta can be represented as  $k^{\mu}_i=(z_i P^{+},k_i^-,\underline{k}_i)$, with $z_i$ being the fraction of the initial $P^+$ momentum which is carried by the gluon labeled by $i$ and $\uvec{k}_i$ being the transverse component of the gluon momentum.
In the 
 LFPT \cite{Weinberg:1966jm,Bjorken:1970ah,Lepage:1980fj,Brodsky:1997de}  one has  to evaluate the energy denominators for each of the intermediate states for the process depicted in Fig.~\ref{fig:onium_n}.  The energy denominator for the $n$ gluons in the intermediate state  is defined to be a difference between the light-front energies of the initial and the  intermediate state in question. For the wave function shown in Fig.~\ref{fig:onium_n} we assume  the last intermediate state is with $n$ gluons. The corresponding energy denominator for this state  reads

\begin{equation}
\overline{D}_n= P^{-}-\sum_{i=1}^n k_i^-=- \frac{1}{P^+}\left({Q^2 \over z_0}-{\uvec{q}^2 \over z_0}+ {\uvec{k}_1^2 \over z_1} + 
{\uvec{k}_2^2 \over  z_2} + \ldots +
 {\uvec{k}_n^2 \over z_n}\right) = - \frac{1}{P^+} D_n\; ,
\label{eq:denominator_n}
\end{equation}
where the light-front energy of the gluon $i$ is equal to $$k_i^- = \frac{\underline{k}_i^2}{ z_i P^+} \; ,$$ and we introduced the auxiliary notation for the (rescaled) denominator $D_n$.

The  wave function shown in example in Fig.~\ref{fig:onium_n} would thus
be given schematically by the expression
$$
\psi_n \sim g^n \Pi_{j=1}^n \frac{V_j}{z_j D_j} \;, 
$$
where $V_j$ are the vertices and $z_j$ and $D_j$ are the corresponding fractional momenta and denominators for all the intermediate states.

The results derived in \cite{ms} and in the following sections are for the color ordered multi-gluon amplitudes. Therefore we focus only on the kinematical parts of the subamplitudes.

In Ref.~\cite{ms} only a special component of the gluon wave function was considered which lead to many simplifications and allowed for the recursion formulae to be solved relatively easily. Namely, it was assumed that the gluons have all the same (e.g. positive) helicites. The convention used in this work is such that the leftmost gluons have incoming momenta and rightmost gluons have outgoing momenta and the light-front time on figures flows from left to right. 
Following \cite{ms} we choose to work 
 in the light-cone gauge, $\eta\cdot A = 0$, with vector
$\eta^{\;\mu} = (0,2,\uvec{0})$ in the light-front coordinates. 
The polarization four-vectors of the gluon with four-momentum $k$ have the following form in this gauge
\be
\epsilon^{(\pm)} = \epsilon_{\perp} ^{(\pm)} + 
{2\uvec{\epsilon}^{(\pm)}\cdot \uvec{k} \over \eta\cdot k}\, \eta\; ,
\ee
where
$\epsilon_{\perp} ^{(\pm)} = (0,0,\uvec{\epsilon} ^{(\pm)})$, and
the transverse vector  is defined by 
$\uvec{\epsilon}^{(\pm)} = \mp {1\over \sqrt{2}}(1,\pm i)$.
Due to the intrinsic properties of the three and four gluon vertices when projected onto helicity states, the resulting gluon wave function with the same (positive) gluon helicities in the final step can be formed with the gluons with positive helicities at all intermediate steps. The 4-gluon vertex vanishes for this helicity configuration and therefore the splittings are given only by the 3-gluon vertex 
with a  $(+ \to ++)$ helicity projection, which corresponds to the situation depicted in Fig.~\ref{fig:onium_n}. 
For the case of interest the projection of the vertex reads
\be
\tilde V_{-++} ^{a_1 a_2 a_3} (k_1,k_2,k_3) 
= 2\, g\,\delta(z_1+z_2+z_3)\,\delta^{(2)}(\uvec{k}_1+\uvec{k}_2+\uvec{k}_3) \,f^{a_1 a_2 a_3} \, z_1 \,
\uvec{\epsilon}^{(+)} \cdot
\left( {\uvec{k}_3 \over z_3} - {\uvec{k}_2 \over z_2}\right) \; .
\label{eq:vhel}
\ee 

Here $g$ is the strong coupling constant and $f^{a_1 a_2 a_3}$ is the structure constant for the $SU(N_c)$ color group.  In the above notation we assumed all gluons to be outgoing, therefore the gluon $1$ has a negative helicity which translates onto positive one when inverting the momentum flow.
For the case of interest, namely that of $(+\to ++)$ transition, the amplitude is described 
by Eq.~(\ref{eq:vhel}), with $z_1$ being the fraction of the $+$ component  of the 
momentum of  the incoming gluon. Then the  vertex is proportional to the   variable 
\be
\uvec{v}_{32} \; \equiv \; \left( {\uvec{k}_3 \over z_3} - {\uvec{k}_2 \over z_2}\right).
\label{eq:v23}
\ee
It is well known \cite{Susskind:1967rg,Kogut:1969xa,Bjorken:1970ah} that on on the light-front the Poincar\'e group can be decomposed onto a subgroup which contains the Galilean-like nonrelativistic dynamics in 2-dimensions. The '$+$' components of the momenta can be interpreted as the 'masses'. In this case
the variable (\ref{eq:v23}) can be interpreted as a relative transverse light-front velocity of the two gluons.  
Interestingly enough, the same variable is present when we consider the change of the energy denominator due to
the splitting.  The importance of this variable is evident when considering the explicit forms of the energy denominators.  For example for the situation depicted in Fig.~\ref{fig:onium_n}
the difference in the energy denominators for the last and last-but-one intermediate state can be shown to be equal
\be
D_{n} - D_{n-1} = {z_3 z_4 \over z_3 + z_4} \, 
\left( {\uvec{k}_3 \over z_3} - {\uvec{k}_4 \over z_4}\right)^2 \, = \,  \xi_{34}\, \uvec{v}_{34}^2\; .
\label{eq:dendifference}
\ee
Here we introduced another combined variable
\be
\xi_{ij} \equiv {z_i z_j \over z_i + z_j} \; ,
\ee 
which depends only on the longitudinal degrees of freedom. It can be interpreted as the reduced mass for a two body problem \cite{Kogut:1969xa}. Thus the difference in the energy denominators  before and after the splitting is proportional to the (twice) kinetic energy  associated with the relative motion of particles $3$ and $4$.

In the previous work we have adopted the convention where the normalization factors for the intermediate lines were absorbed into the vertices, i.e.
\be
\bar V_{\lambda_1 \lambda_2 \lambda_3} ^{a_1 a_2 a_3} (k_1,k_2,k_3) = 
{1\over \sqrt {z_1 z_2 z_3}} \, \tilde V_{\lambda_1 \lambda_2 \lambda_3} ^{a_1 a_2 a_3} (k_1,k_2,k_3)
= 2 g f^{a_1 a_2 a_3} \,
{\uvec{\epsilon}^{(+)} \uvec{v}_{32} \over  \sqrt{\xi_{32}}} \; .
\label{eq:vnorm}
\ee

Using the definitions of the vertex and the energy denominators presented above one can form the recurrence formula for the gluon wave function with $n$ components
\be
\Psi_{n} (k_1,\ldots, k_i, k_{i+1}, \ldots, k_{n}) = 
{2 ig \over \sqrt{\xi_{i\, i+1}}} 
{\uvec{\epsilon}^{(+)} \uvec{v}_{i\, i+1} \over  D_{n-1} + \xi_{i\, i+1}\,
\uvec{v}_{i\, i+1}^2}\, \Psi_{n-1} (k_1, \ldots, k_{i\,i+1},\ldots, k_{n})\; .
\label{eq:vxisplit}
\ee 
Here, 
 $\Psi_{n-1}(k_{1},k_2,\ldots, k_{i\,i+1}, \ldots, k_{n})$ (with $k_{i\,i+1}\equiv k_i+k_{i+1}$) is the $(n-1)$-gluon wave function in momentum space before the splitting of gluon with momentum $k_{i\, i+1}$, and  
$\Psi_{n} (k_1,k_2,\ldots, k_i,k_{i+1}, \ldots, k_{n})$ is the wave function after 
splitting of this gluon.  To obtain the full wave function one   
 needs to sum over the different possibilities of the splittings which  gives the following result
\be
\Psi_{n} (k_1,k_2,\ldots, k_{n}) = 
{2 ig \over D_{n}}\, \sum_{i=2}^{n} 
{\uvec{\epsilon}^{(+)} \uvec{v}_{i-1 \, i} \over  \sqrt{\xi_{i-1 \, i}} }\, \Psi_{n-1} (k_1,\ldots ,k_{i-1 \, i},\ldots, k_{n})\; ,
\label{eq:recurrence1}
\ee
where $D_{n}$ is the denominator for the last of all the intermediate states with $n$ gluons. 
We note that all dependence on momenta of daughter gluons $i$~and~$i-1$ 
is now encoded in two variables: $\xi_{i-1 \, i}$ and $\uvec{v}_{i-1 \, i}$. 
The recurrence formula will be utilized for the derivation of the scattering amplitudes. We note that this recurrence formula is written for the special choice of the helicities. It can be readily generalized for the other case of helicities. We will come back to this point later when we use it to derive the Parke-Taylor amplitude.


The variables  $\underline{v}_{jk}$  which represent the 'velocities' of gluons on the light-front   are actually related to the variables 
used in the framework of helicity amplitudes, see \cite{Mangano:1990by}
for a nice review. 
For a  given pair of on-shell momenta $k_i$ and $k_j$ we have the result  
\be
\langle ij \rangle = \sqrt{z_i z_j} \; \uvec{\epsilon}^{(-)} 
\cdot \left( {\uvec{k}_i \over z_i} -  {\uvec{k}_j \over z_j} \right)\, ,
\qquad 
[ij] =  \sqrt{z_i z_j} \; \uvec{\epsilon}^{(+)}\cdot 
\left( {\uvec{k}_i \over z_i} -  {\uvec{k}_j \over z_j} \right)\, ,
\label{eq:ij}
\ee
where the latter are defined by
\be
\langle i  j \rangle = \langle i- | j+ \rangle\, ,  \; \; \; [ij] = \langle i+ | j- \rangle \; .
\label{eq:ijdef1}
\ee
The  chiral projections of the spinors for massless particles  are defined as
\be
|i \pm \rangle \; = \; \psi_{\pm}(k_i) \; = \; \frac{1}{2}(1\pm\gamma_5)\psi(k_i) \; \; , \;\;\;\;\langle \pm i| \; = \; \overline{\psi_{\pm}(k_i)} \; ,
\label{eq:ijdef2}
\ee
for a given momentum $k_i$.
The spinor products are complex square roots of the total energy mass squared for the pair of gluons $(i,j)$
\be
\langle \,ij\, \rangle [\,ij\,] = (k_i + k_j)^2 ,
\label{eq:ijs}
\ee
and they also satisfy $\langle ij \rangle = [ij]^*$.
Using the above definitions (\ref{eq:ij},\ref{eq:ijs}) we have that
$$
\langle \,ij\, \rangle [\,ij\,]=  z_i z_j \, \left( {\uvec{k}_i \over z_i} -  {\uvec{k}_j \over z_j}  \right)^2 \; .
$$
which is real and positive for the on-shell gluon momenta.
Finally, combining (\ref{eq:v23}) and (\ref{eq:ij}) we obtain
\be
\langle\, ij\,\rangle = 
\sqrt{z_i z_j} \; \uvec{\epsilon}^{(-)}\cdot \uvec{v}_{ij}\, ,
\qquad 
[\,ij\,] = \sqrt{z_i z_j} \; \uvec{\epsilon}^{(+)}\cdot \uvec{v}_{ij} \, ,
\ee
and the dependence on the transverse momenta in the light-front wave function
can be expressed by spinor products $\langle\,ij\,\rangle$ and $[\,ij\,]$.

\subsection{The multi-gluon wave function in the case of the on-shell incoming gluon}

The multi-gluon wave function $\Psi_n$ has a closed compact form in the case 
of an on-shell incoming gluon, $Q^2=0$ with helicity~$+$ and when all the final gluons also have $+$ helicities.  The simplification stems from the fact that the   energy denominators  do not contain the virtuality $Q^2$. 

\begin{figure}[h]
\centering
\subfloat[]{\includegraphics[width=.2\textwidth]{{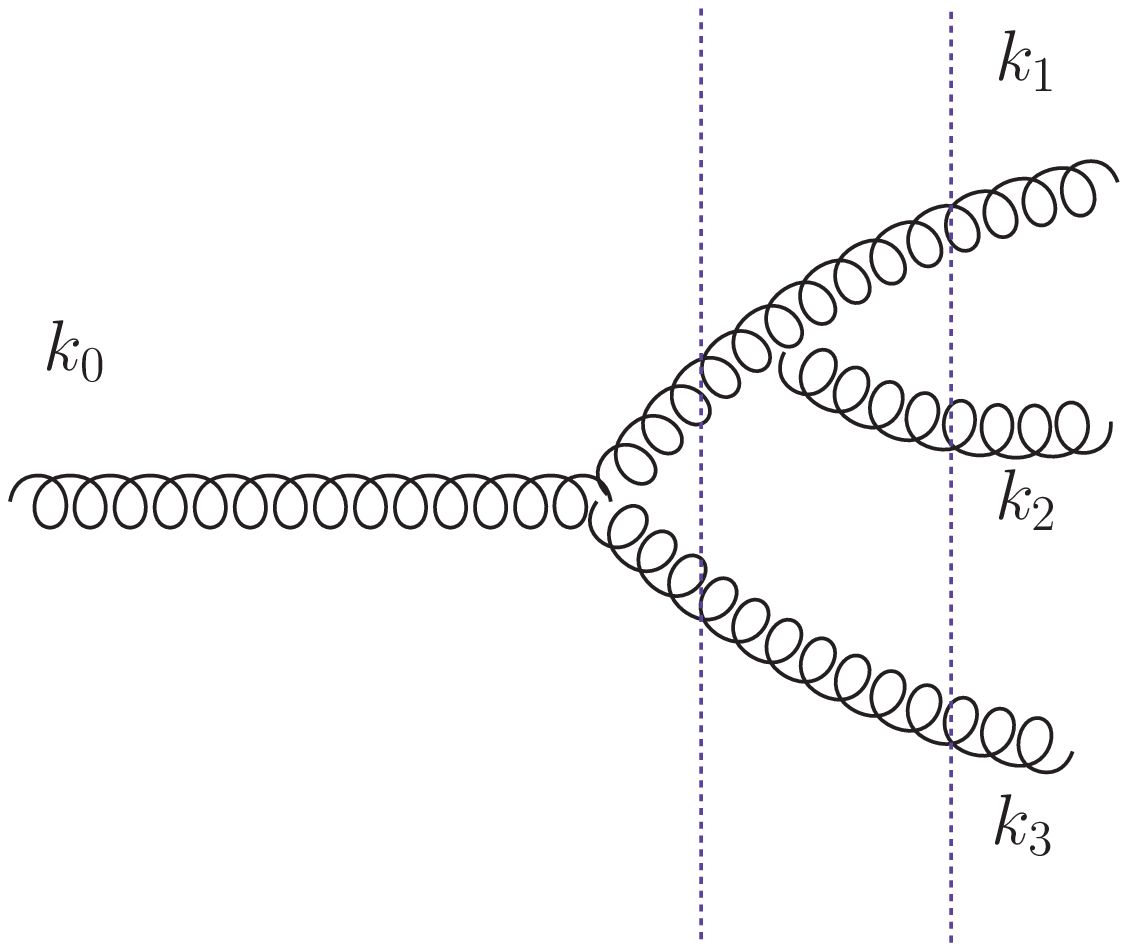}}}\hspace*{2cm}
\subfloat[]{\includegraphics[width=.2\textwidth]{{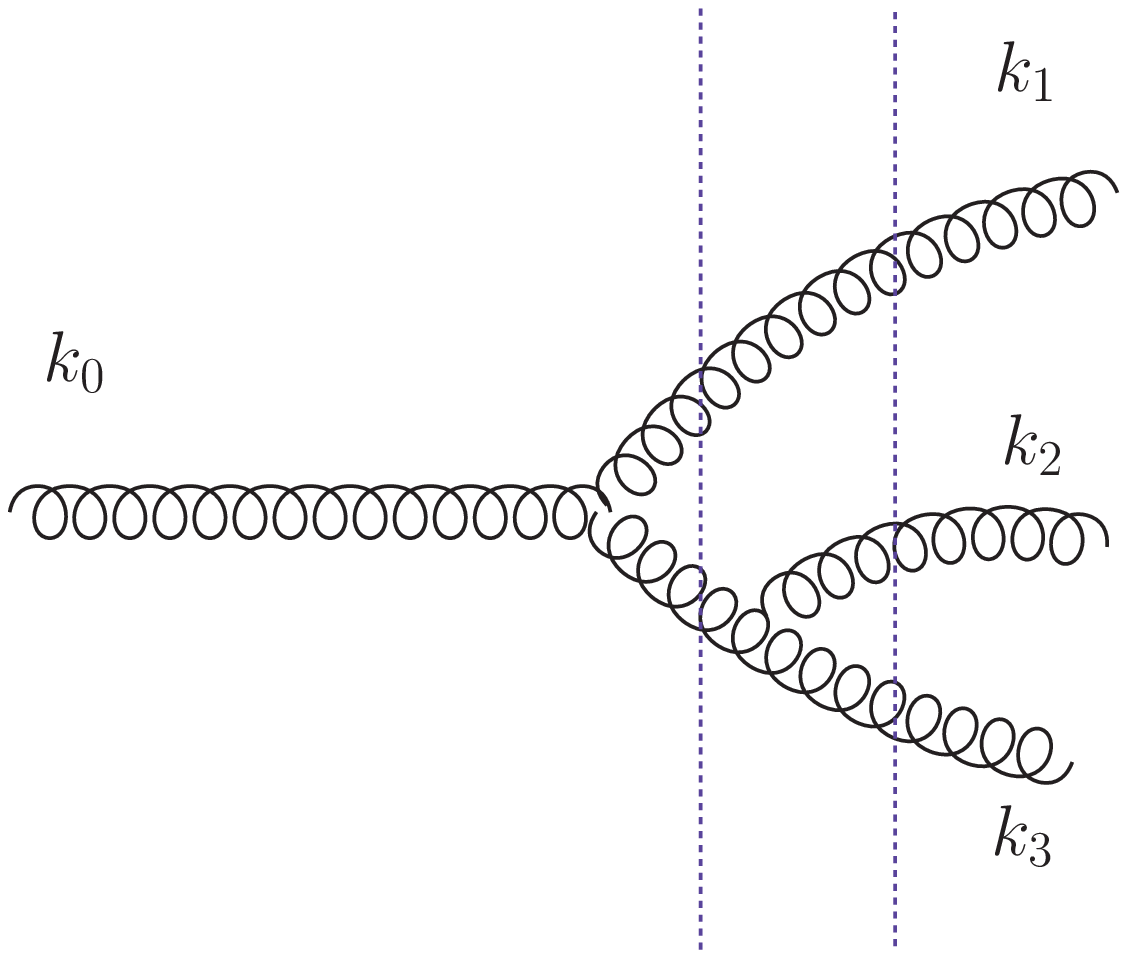}}}
\caption{Graphs contributing to the gluon wave function with 3 gluons.}
\label{fig:wf3}
\end{figure}

The recurrence formula, i.e. Eq.~(\ref{eq:recurrence1}), for the case of the on-shell incoming gluon and the same helicities for the outgoing gluons has the explicit form

\[
D_{n+1}\, \Psi_{n+1}(1,2,\ldots, n+1) \; = \; 
\]
\be
= \,2ig{v^*_{12} \over \sqrt{\xi_{12}}} \Psi_n (12,3,\ldots,n+1) \, +
 \,2ig{v^*_{23} \over \sqrt{\xi_{23}}} \Psi_n (1,23,\ldots,n+1) \, +
  \, \ldots \, +2ig{v^*_{n\, n+1} \over \sqrt{\xi_{n\,n+1}}} \Psi_n (1,2,\ldots,n\, n+1)\, ,
\label{eq:fullsplit}
\ee
with $D_{n+1}=\uvec{k}_1^2/z_1+\uvec{k}_2^2/z_2+\dots+\uvec{k}_{n+1}^2/z_{n+1}-\uvec{q}^2/z_0$.
We have introduced the notation $\Psi_n(1,\ldots,i-1 \, i,\ldots,n+1)$ where $ \; i-1 \, i \; $ means that it is the gluon with  the transverse momentum $k_{i-1\, i}=k_{i-1}+k_i$.
Again, we recall that we are considering  color ordering in the amplitudes, therefore we will suppress color degrees of freedom.
The  wave function for   the incoming state has the  normalization
$
\Psi_1(1) = 1
$.
After the first splitting one gets
\be
\Psi_2(1,2) \; =  \;
2ig\, {1 \over \sqrt{\xi_{12}}}\,
{v^*_{12} \over \xi_{12} \uvec{v}_{12}^2} 
\; = \;  
-ig\, {1\over\sqrt{\xi_{12}}}\, {1\over \xi_{12} v_{12}} \, ,
\ee
where we have taken that $Q^2=0$.
According to (\ref{eq:fullsplit}), 
the next splitting leads from $\Psi_2(1,2)$ to $\Psi_3(1,2,3)$ 
with the graphs depicted in
Fig.~\ref{fig:wf3} and the result is 

\[
D_3 \Psi_3(1,2,3) \; = \;2i g\, \left[ \,
{v^*_{12} \over \sqrt{\xi_{12}}}\Psi_2 (12,3) 
\, + \, {v^*_{23} \over \sqrt{\xi_{23}}} \Psi_2 (1,23) \, \right]
\]
\be
= \;2 g^2\, \left[ \, {v^*_{12} \over \sqrt{\xi_{12}\xi_{(12)3}}}
{1\over\xi_{(12)3}\, v_{(12)3}} 
\,+\,{v^*_{23} \over \sqrt{\xi_{23}\xi_{1(23)}}}
{1\over\xi_{1(23)}\, v_{1(23)}} \, \right] = 2 g^2{\sqrt{z_0} \over \sqrt{z_1 z_2 z_3}}\,
{v^*_{12} \xi_{1(23)} v_{1(23)} + v^*_{23}\xi_{(12)3}v_{(12)3} \over\xi_{(12)3}\xi_{1(23)} v_{(12)3} \, v_{1(23)}} \, ,
\label{eq:wavefunction3}
\ee
where we used 
$\xi_{12}\xi_{(12)3} \, = \, \xi_{23}\xi_{1(23)} \, = \, 
{z_1 z_2 z_3 \over z_1 + z_2 + z_3} \; = \; {z_1 z_2 z_3 \over z_0}$.
We can further simplify the numerator of this expression to  obtain,
\be
\Psi_3(1,2,3) \; = \;- g^2 \, {\sqrt{z_0} \over \sqrt{z_1 z_2 z_3}}\,
{1\over\xi_{(12)3}\xi_{1(23)}}\, {1\over v_{(12)3} \, v_{1(23)}}\; . 
\label{eq:wavefunction3simple}
\ee 
Note that, the energy denominator $D_3$ disappeared from the equation as it has canceled with the numerator when finding the common denominator for expression (\ref{eq:wavefunction3}).
This is actually quite an important property and leads to the very simple expression for the resummed gluon wave function. It is also important in recovering some crucial  properties of the helicity amplitudes as we shall show later.
The resummed wave function with $n$-gluon components can be showed to be given by  the following compact expression \cite{ms}
\[
\Psi_n(1,2,\ldots,n) \; = \; (-1)^{n-1}(ig)^{n-1}\,\Delta^{(n)}\, {\sqrt{z_0} \over \sqrt{z_1 z_2 \ldots z_n}}\,
{1\over\xi_{(12\ldots n-1)n}\,\xi_{(12\ldots n-2)(n-1\,n )} \,
\ldots \, \xi_{1(2\ldots n)}}
\]
\be
\times\;
{1\over v_{(12\ldots n-1)n}\, v_{(12\ldots n-2)(n-1\,n )} \,
\ldots \, v_{1(2\ldots n)}} \; . 
\label{eq:psinfact}
\ee
Throughout this section we have been using the following complex representation of the transverse vectors: 
$v_{ij} = \uvec\epsilon^{(-)}\cdot \uvec{v}_{ij}$,  
$v^*_{ij} = \uvec\epsilon^{(+)}\cdot \uvec{v}_{ij}$, and 
a useful notation, 
\be
{v}_{(i_1 i_2 \ldots i_p)(j_1 j_2 \ldots j_q)} = 
{{k}_{i_1} + {k}_{i_2} + \ldots + {k}_{i_p}
\over z_{i_1} + z_{i_2} + \ldots + z_{i_p}} -
{{k}_{j_1} + {k}_{j_2} + \ldots + {k}_{j_q}
\over z_{j_1}+ z_{j_2} + \ldots + z_{j_q}}\; ,
\ee
\be 
\xi_{(i_1 i_2 \ldots i_p)(j_1 j_2 \ldots j_q)} = 
{
(z_{i_1}+z_{i_2}+\ldots+z_{i_p})
(z_{j_1}+z_{j_2}+\ldots+z_{j_q})
\over 
z_{i_1}+ z_{i_2}+ \ldots+ z_{i_p} + z_{j_1} + z_{j_2} + \ldots + z_{j_q}} \;,
\ee
together with  $k_i \equiv \uvec{\epsilon}^{(+)}\cdot \uvec{k}_i$. We also introduced notation for the partial sums 
$z_{(1\dots i)}\equiv z_1+z_2+\dots+z_i$ and
$k_{(1\dots i)}\equiv k_1+k_2+\dots+k_i$.
In formula (\ref{eq:psinfact}),  $\Delta^{(n)}$ is defined to contain  the  global momentum conservation $\delta$-functions, 
$\Delta^{(n)} \equiv \delta^{(2)} \left(\, \sum_{i=1} ^n \uvec{k}_i \, -\uvec{q} \right) \, 
\delta \left(\, 1 - \sum_{i=1} ^n z_i\, \right)$. Recall, that on the light-front, the longitudinal momenta and the transverse momenta are conserved but not the light-front energies $k^-$ (except for final and initial states).

The formula (\ref{eq:psinfact}) can be proven by induction (for details see \cite{ms}) and with the help of  the following identity
\be
v_{(1\ldots i)(i+1 \ldots n)} \xi _{(1\ldots i)(i+1 \ldots n)}
\; = 
 \sum_{j=1}^{i} k_j \, - \, \frac{z_{(1\dots i)}}{z_{(1\dots n)}}\, \sum_{j=1}^{n} k_j \, = \,  k_{(1\ldots i)} -\frac{z_{(1\dots i)}}{z_{(1\dots n)}}k_{(1\ldots n)}=k_{(1\ldots i)} -{z_{(1\dots i)} \over z_0} q\;,
 \label{eq:vxi_sumk_gen}
\ee
where the momentum conservation conditions $q=k_{(1\ldots n)}$ and $z_0=z_{(1\ldots n)}$ have been utilized.
The above identity can be used to show that 
\begin{equation}
\sum_{i=1}^{n-1} v^*_{i\,i+1}\xi_{(1\dots i)(i+1\dots n)} v_{(1\dots i)(i+1\dots n)} \; =
\; -{1 \over 2} \left( \sum_{i=1}^{n} \frac{\uvec{k}_i^2}{z_i} -{\uvec{q}^2 \over z_0} \right)\; = \; -{1 \over 2} D_{n} \; .
 \label{eq:sumvxi}
\end{equation}

The above expression is what appears in the numerator after combining the different terms
in the recursion formula to a common denominator. It therefore cancels with an overall denominator for the wave function leading to the simple form of the wave function (\ref{eq:psinfact}). The example of that property was used to simplify  Eq.~(\ref{eq:wavefunction3}).

\subsection{Fragmentation of an off-shell  gluon}
\label{sec:fragm}

\begin{figure}[ht]
\centerline{\includegraphics[width=0.4\textwidth]{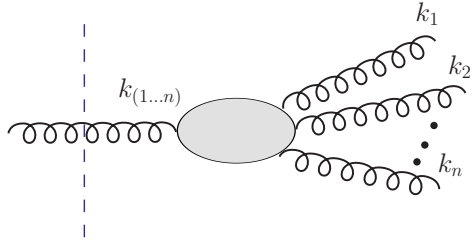}}
\caption{Pictorial representation of the fragmentation amplitude $T[(12\ldots n) \to 1,2,\ldots, n]$
for the single off-shell initial gluon.}
\label{fig:fragsinglegluon}
\end{figure}

Using  the techniques developed for the wave functions one can similarly resum the graphs in a kinematical situation when the initial state is an off-shell gluon, and then it fragments to give the final $n$-gluons which are on-shell. There are interesting similarities between the two cases, i.e. wave functions and fragmentation functions, which can be expected as the topology of the graphs is exactly the same. There are some important differences however, which are related to the properties of factorization. Namely, the fragmentation functions can be shown to factorize whereas the gluon wave functions do not exhibit this property (at least not in momentum space).
Using the notation of the previous section one can show that for the case of the splitting of an off-shell gluon, denoted by (12),  into 2 on-shell gluons (denoted respectively by 1 and 2) the fragmentation amplitude can be expressed as

\be
T_2[(12) \to 1,2]  \; = \;  
-\frac{2ig}{\tilde{D}_2}  {v^*_{12} \over \sqrt{\xi_{12}} } =
-2ig \, {v^*_{12} \over \xi^{3/2} _{12} |\uvec{v}_{12}|^2}\,  = ig \, \xi_{12}^{-3/2} \, \frac{1}{v_{12}} \; ,  
\ee
where $\tilde{D}_2=\sum_{i=1}^2 {\uvec{k}^2_i / z_i} - \left(\uvec{k}_{(12)}\right)^2 / z_{(12)}$ for this case.  The general expression  for the fragmentation part of the amplitude for 1 to n gluons as $T_n[(12\ldots n) \to 1,2,\ldots, n]$ on  a tree-level is given by \cite{ms}
\be
\label{eq:frag1}
T_n[(12\ldots n) \to 1,2,\ldots, n] \; = \; 
(ig)^{n-1} \left( {{z_{(12\ldots n)}} \over {z_1 z_2 \ldots z_n}} \right)^{3/2} \;
{1\over v_{12} v_{23} \ldots v_{n-1\, n}} \; .
\ee
The off-shell gluon is labeled by $(12\ldots n)$  and the final state
 $n$ on-shell gluons are labeled as $1,2,\ldots n$.
These $n$ final state gluons have transverse momenta  $\uvec{k}_1,\ldots,\uvec{k}_n$ and
the longitudinal momentum fractions $z_1,\ldots,z_n$. The initial gluon has transverse momentum $\uvec{k}_{(1\ldots n)}$ and the longitudinal fraction $z_{(1\ldots n)}$ where we again used shortcut notation $\uvec{k}_{(1\ldots n)}=\sum_{j=1}^{n} \uvec{k}_j$ and   $z_{(12\ldots n)} = \sum_{i=1} ^n z_i$ .
The fragmentation function is depicted in Fig.~\ref{fig:fragsinglegluon}.

The above formula can be again proven by mathematical induction using analogous relations to the  (\ref{eq:vxi_sumk_gen}) and (\ref{eq:sumvxi}). On top of that,
one crucial assumption is that of the factorization property of the fragmentation functions
which are formed from topologically disconnected trees and originate from different off-shell parents.
Physically, the factorization property of independent tree amplitudes should be quite 
intuitive. One considers a  fragmentation tree with a topology 
$\Theta = \Theta_1 \cup \ldots \cup \Theta_m$ where 
$\Theta_1,\ldots \Theta_m$ are topologies of the fragmentation trees 
of $m$~parent virtual gluons into $n$~on-shell gluons,
so that the first parent gluon, denoted by $(12\ldots n_1)$ fragments 
into gluons $(1,2,\ldots, n_1)$, the second parent gluon,  
$(n_1+1, n_1+2,\ldots, n_1)$ fragments into gluons $(n_1+1,n_1+2,\ldots, n_2)$,
and so on.
One can then show that the amplitude of the fragmentation tree $\Theta$, denoted by
$T_{\Theta}$ may be factorized into fragmentation amplitudes of  
the parent gluons, $T_{\Theta_i}$ in the following way,
$T_{\Theta}[(1 \ldots n_1),(n_1+1\ldots n_2),\ldots,
(n_{m-1}\ldots n) \, \to \, 1,2,\ldots n]$, 
\begin{multline}
\label{eq:factm}
T_{\Theta}[(1 \ldots n_1),(n_1+1\ldots n_2),\ldots,
(n_{m-1}\ldots n) \, \to \, 1,2,\ldots n]
\; = \; 
T_{\Theta_1}[(1 \ldots n_1)\, \to \, 1,2,\ldots n_1]\, 
\\
\times \;
T_{\Theta_2}[(n_1+1 \ldots n_2)\, \to \, n_1+1,n_1+2,\ldots n_2]\,\times
\ldots\, \times\,
T_{\Theta_m}[(n_{m-1}+1 \ldots n)\, \to \, n_{m-1},n_{m-1}+1,\ldots n] \; .
\end{multline}
The factorization property is trivial in the case of standard time-ordered perturbation theory.  However, it is 
 not obvious in the light-front formulation, as for
a given time-ordering of the splittings, the variables related to 
different trees are correlated in the energy denominators, and the factorization
property  does not hold for an individual diagram with particular ordering of the vertices on the light-front. It only holds if the 
complete summation over all possible  orderings of the light-front time-orderings is performed within all connected trees, while preserving the topologies as has been demonstrated in \cite{ms}.

\begin{figure}[ht]
\centerline{\includegraphics[width=0.5\textwidth]{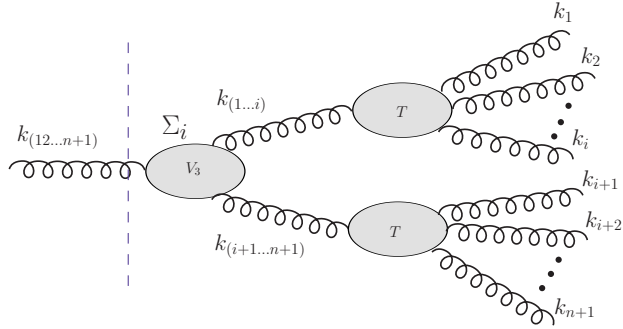}}
\caption{Pictorial representation of the factorization property represented in Eq.~(\ref{eq:fragonestep}), a light-front analog of the Berends-Giele recursion relation. The helicities of the outgoing gluons are chosen to be the same in this particular case. Dashed vertical line indicates the energy denominator $\tilde{D}_{n+1}$.}
\label{fig:fragmaster}
\end{figure}

The above factorization property can be used then to write down the explicit recursion formula for the fragmentation amplitudes.
Namely, the fragmentation into $n+1$ gluons denoted by $T_{n+1}[(1,2,\ldots, n+1) \to 1,2,\ldots, n+1]$ 
can be represented by lower fragmentation factors  
$T_i[ (1\ldots i)  \to 1,\ldots,i\,]$ and 
$T_{n+1-i}[ (i+1\ldots n+1)\to i+1,\ldots,n+1]$  and by summing over the 
splitting combinations. To be precise one has
\begin{multline}
T_{n+1}[(12 \ldots n+1) \to 1,2,\ldots,n+1] \; = \; -{2ig\over \tilde D_{n+1}} \; 
\sum _{i=1} ^n  \, \left\{\,
{v^*_{(1\ldots i)(i+1\ldots n+1)} 
\over \sqrt{\xi_{(1\ldots i)(i+1\ldots n+1)}}} \; \right. \\
\left. \times \;\rule{0em}{1.8em}
T_i[(1\ldots i) \to 1,\ldots,i\,] \; T_{n+1-i}[(i+1\ldots n+1) \to i+1,\ldots,n+1]\,
\right\}\, . 
\label{eq:fragonestep}
\end{multline}
This expression is the final state analog of formula (\ref{eq:recurrence1}) for the iteration of the wave function and it is schematically depicted in Fig.~\ref{fig:fragmaster}.

The above defined fragmentation functions $T$ can be related to the gluonic currents which play an important role in the Berends-Giele recursion relations.
These recursive relations utilize the (gauge-dependent) current $J^{\mu}$, which is obtained from the amplitudes by taking  one of the gluons off-shell. The dual subamplitudes can be obtained by contraction with the polarization vector and putting the gluon back on-shell, \cite{Mangano:1990by}
\be
M(0,1,2,\dots,n) = i P^2\epsilon^{\mu} J_{\mu}(1,2,\dots,n)_{P=-P_0}
\ee
where we have defined $P=\sum_{i=1}^n P_n$ and in this formula $p_i$ denote the four-vectors for the momenta of the outgoing particles.
In the case of the light-front calculation the light-front current can also be defined and related to the fragmentation function in the following way
$$
T_n((1,2,\dots,n)\rightarrow 1,2,\dots,n) \equiv \epsilon^{\mu}(12\dots n) J_\mu (1,2,\dots,n)
$$
where by $\epsilon^{\mu}(12\dots n)$ we denote the polarization vector of the 
incoming (off-shell) gluon in the fragmentation function.
With such definition the factorization property for the fragmentation function \eqref{eq:fragonestep} is a light-front analog of the Berends-Giele \cite{Berends:1987me} recursion formula (see also \cite{Kosower:1989xy} for the recurrence relations in the light-cone gauge)
\begin{multline}
J^{\mu}(1,2,\dots,n) = -\frac{i}{P^2}\sum_{i=1}^{n-1} V_3^{\mu \nu \lambda}(p_{1\dots i},p_{i+1\dots n}) J_{\nu}(1,\dots,i) J_{\lambda}(i+1,\dots,n)\\
-\frac{i}{P^2} \sum_{i=j+1}^{n-1} \sum_{j=1}^{n-2} V_4^{\mu \nu \lambda \delta} J_{\nu}(1,\dots,j) J_{\lambda}(j+1,\dots,i) J_{\delta}(i,\dots,n) \; .
\label{eq:bg}
\end{multline}
The simpler form of \eqref{eq:fragonestep} (as compared to \eqref{eq:bg}) which only includes three gluon vertex stems from the fact that it has been written for particular configuration of helicities. It is possible to write down a general factorization (recursion) relation for the fragmentation function which will include the 4-gluon vertex as well as the Coulomb term.
For example in the case of the fragmentation into $n-1$ gluons with positive helicity and $1$ gluon with negative helicity the recursion relation of the fragmentation function has the following form
\begin{multline}
T_{n+1}[(12 \ldots n+1)^+ \to 1^-,2^+,\ldots,n+1^+] \; = \; -{2ig\over \tilde D_{n+1}} \; 
 \left\{\, \sum _{i=2} ^n
{v^*_{(1\ldots i)(i+1\ldots n+1)} 
\over \sqrt{\xi_{(1\ldots i)(i+1\ldots n+1)}}} \; \right. \\
\left. \times \;\rule{0em}{1.8em}
T_i[(1\ldots i)^+ \to 1^-,2^+,\ldots,i^+\,] \; T_{n+1-i}[(i+1\ldots n+1)^+ \to i+1^+,\ldots,n+1^+] \right. \\
\left. +\sum _{i=1} ^n v_{(1\ldots i)(1\ldots n+1)} 
\sqrt{{z_{i+1\ldots n+1}} \over {z_{1\ldots i}z_{1\ldots n+1}}} \;  
T_i[(1\ldots i)^- \to 1^-,2^+,\ldots,i^+\,] \; T_{n+1-i}[(i+1\ldots n+1)^+ \to i+1^+,\ldots,n+1^+] \right\} \\
+{i\over \tilde D_{n+1}}\sum _{i=1} ^{n-1} \sum_{j=i+1}^n V_4^{-++} \;  
T_i[(1\ldots i)^- \to 1^-,2^+,\ldots,i^+\,] \; T_{j-i}[(i+1\ldots j)^+ \to i+1^+,\ldots,j^+]T_{n+1-j}[(j+1\ldots n+1)^+ \to j+1^+,\ldots,n+1^+]  .\\ 
\label{eq:fragonesteponeg}
\end{multline}
In the above formula the vertex $V_4^{-++}$ contains both 4-gluon vertex and the Coulomb term
\be
V_4^{-++}= {2ig^2 \over \sqrt{z_{1\dots n+1}\;z_{1\ldots i}\;z_{i+1\ldots j}\;z_{j+1\ldots n+1}}} {z_{1\ldots n+1}\;z_{i+1\ldots j}-z_{1\ldots i}\;z_{j+1\ldots n+1} \over z_{1\ldots j}^2}
\ee
As mentioned above the Berends-Giele relations can be written on the level of individual diagrams, whereas for the derivation of the analogous recursion relations on the light-front
\eqref{eq:fragonestep}, \eqref{eq:fragonesteponeg} the summation over the time-ordering is necessary to decouple the disconnected fragmentation trees.

Thus using wave functions and fragmentation functions one can write two distinct recursion relation in the LFPT. There is however an important difference between the recurrence formulae in the two cases on the light-front, i.e. fragmentation and wave functions. The factorization property enjoyed by  the fragmentation amplitudes does not hold in the case of the wave functions. This stems from the fact that in the latter case one  still has the energy denominator corresponding to the $n$  gluons, see Fig.~\ref{fig:onium_n}. This means that the $n$  gluons of the wave function are in the intermediate state and still have to interact. Therefore they do not necessarily form disconnected trees and therefore can be correlated and cannot be  factorized. Still, the recursion formulae for the gluon wave functions will prove to be  very useful, and can be used to obtain the helicity amplitudes   as we will see in the following sections.


\section{MHV scattering amplitudes from the light-front wave function}
\label{sec:mhv}

In the previous sections we have analyzed the (tree-level) wave function with arbitrary number of gluons in the light-front formulation as  well as the fragmentation function of the off-shell gluon into $n$ final state gluons.
The main goal of this section is to show how to reproduce on the light-front the helicity amplitudes
from the previously derived wave functions. We recall that the 
  Parke-Taylor amplitudes \cite{Parke:1986gb} (see \cite{Mangano:1990by} for a comprehensive review) have the following form
\be
{\cal M}_n \;=\; \sum_{\{1,\dots,n\}} {\rm tr}(t^{a_1}t^{a_2}\dots t^{a_n}) \; m(p_1,\epsilon_1;p_2,\epsilon_2;\ldots;p_n,\epsilon_n) \; ,
\label{eq:ptsum}
\ee
where $a_1,a_2,\ldots,a_n$, $p_1,p_2,\ldots,p_n$ and $\epsilon_1,\epsilon_2,\ldots,\epsilon_n$ are the color indices, momenta and the helicities of the external $n$ gluons. Matrices $t^a$ are in the fundamental representation of the color $SU(N_c)$ group. The sum in (\ref{eq:ptsum}) is over the $(n-1)!$ non-cyclic permutations of the set $\{0,1,\dots,n\}$.

 The kinematical parts of the amplitude denoted by $m(1,2,\ldots,n)\equiv m(p_1,\epsilon_1;p_2,\epsilon_2;\ldots;p_n,\epsilon_n)$ are color independent and gauge invariant and enjoy a number of important properties.
 Among them is the requirement that
 the amplitudes where all the gluons have the same helicities, or only one is different from the others are vanishing (here we used the convention that all the gluons are outgoing) 
$$
m(\pm,\pm,\dots,\pm)=m(\mp,\pm,\pm,\dots,\pm)=0\; .
$$
These properties can be demonstrated  on the tree-level using the supersymmetry relations \cite{Grisaru:1977px}.
The non-vanishing amplitude for the configuration $(-,-,+,\dots,+)$  at the tree-level is given by the formula
\be
m(1^-,2^-,3^+,\ldots,n^+) \, = \, i g^{n-2}\, \frac{{\langle 1 2 \rangle}^4}{\langle1 2  \rangle \langle2 3   \rangle\ldots \langle n\!-\!2\; n\!-\!1   \rangle\langle  n\!-\!1\;n  \rangle \langle n 1 \rangle} \; ,
\label{eq:ptmhv}
\ee
where the spinor products are defined by Eqs.~(\ref{eq:ijdef1}),(\ref{eq:ijdef2}).

In the previous work \cite{ms} a simplified derivation of the Parke-Taylor amplitudes on the light-front has been demonstrated. The simplification originated from the fact that the projectile gluon with the evolved wave function was allowed to scatter off the target gluon which was separated by large rapidity. 
  This allowed the exchange between the projectile and the target to be be treated in the high energy limit.  In this limit the interaction between the projectile and the target is mediated by an instantaneous part of the gluon propagator in the light-cone gauge defined as
$$
D_{\rm inst}^{\mu\nu}  = \frac{\eta^{\mu} \eta^{\nu}}{(k^+)^2}\; .
$$
 The kinematics of the exchange was simplified but the internal structure of the projectile was treated without any simplifications. In addition after the interaction the evolution in the final state was also included. To be precise the fragmentation of the gluons after the instantaneous interaction took place was included through the fragmentation functions $T$ defined above.
 
 After taking this simplification it can be shown that the Parke -Taylor amplitude can be represented in this limit
 in the factorized form 
 
\be
\widetilde M(a,0 \to b,1,2,\ldots n) \, = \, {1\over \sqrt{z_a z_b}}
g^2 {s\over |t|} \;\, \tilde \Psi_n(1,2,\ldots n).
\label{eq:m_psi}
\ee
The factor 
$\sim g^2 s / |t|\,$ comes form the gluon exchange and  $\tilde \Psi_n$ is the amplitude of the gluon evolution of the projectile gluon into $n$-gluon final state.
The factor $1/\sqrt{z_a z_b}$ is present because of the convention. The expression 
for the  general $n$ for the subamplitude $\tilde \Psi_n$ can be found from the following formula
\begin{multline}
\label{eq:psifinmaster}
\tilde \Psi_n(1,2,\ldots ,n) \; = \;
\sum_{m=1} ^n \, \sum_{(1\leq n_1<n_2<\ldots<n_{m-1}\leq n)}\;
\Psi_m((1\ldots n_1)(n_1+1\ldots n_2)\ldots(n_{m-1}+1\ldots n))\\
\times\;
T[(1\ldots n_1)\to 1,\ldots, n_1]\;T[(n_1+1\ldots n_2)\to n_1+1,\ldots, n_2]\,
\ldots \,T[(n_{m-1}+1\ldots n)\to n_{m-1}+1,\ldots, n] \;.
\end{multline}
This formula is obtained by taking into account all possible attachments of the exchanged gluon in the cascade. 
The additional assumption is the 
 factorization of fragmentation of virtual
gluon (\ref{eq:factm}).
  
The final result is the following general form for the $\tilde{\Psi}_n$  for an arbitrary number of emitted gluons
\be
\label{eq:psifin}
\tilde\Psi_n(1,2,\ldots n) \; = \; g^{n-1} \, {k_{(1\ldots n)} \over k_1 / z_1}
\; 
{1 \over \sqrt{z_1 z_2 \ldots z_n}} \; {1 \over z_1 z_2 \ldots z_n} \; 
{1 \over v_{12} v_{23} \ldots v_{n-1\, n}} \; . 
\ee
which can be shown to arise \cite{ms} when the recursion formula \eqref{eq:psifinmaster} is solved using the explicit expressions for the fragmentation functions and the wave functions. Finally it was shown that the (approximated) expression for the Parke-Taylor amplitude can be re-derived in this limit by combining
Eqs.~(\ref{eq:m_psi}) and (\ref{eq:psifin}).

However, as mentioned above this derivation is only approximate since it is based on the assumption of the large rapidity difference between the incoming projectile and the target gluon, which
is only appropriate in the high- energy limit. In the following, we shall show a different method
which will relate the wave functions with the amplitudes and allow to obtain the exact results without any kinematical approximations.

%
\subsection{Simple example: 2 $\to$ 2 amplitude on the light-front}

In order to prepare the ground for the derivation of general relations between gluon wave functions derived earlier and scattering amplitudes, we shall consider at first a lowest order example of the $2\rightarrow 2$
amplitude using the rules of the LFPT and the notation  introduced in the previous sections. The goal of this simple example is to illustrate the fact that the resulting amplitude
 has a very similar form to the gluon wave function for $1 \rightarrow 3$ transition.

\begin{figure}[h]
\centering
\subfloat[]{\label{fig:2to2a}\includegraphics[width=.25\textwidth]{{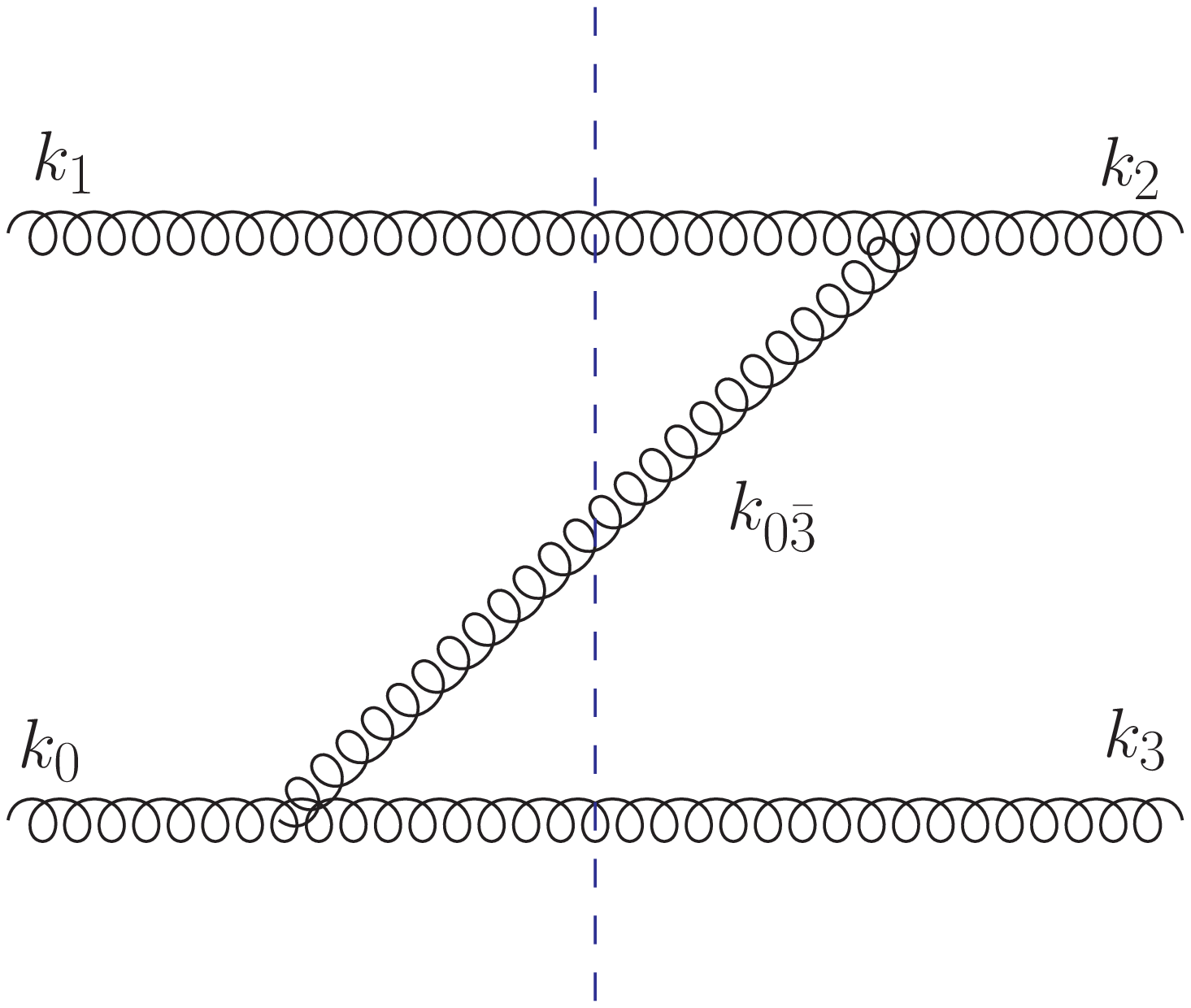}}}\hspace*{0.7cm}
\subfloat[]{\label{fig:2to2b}\includegraphics[width=.25\textwidth]{{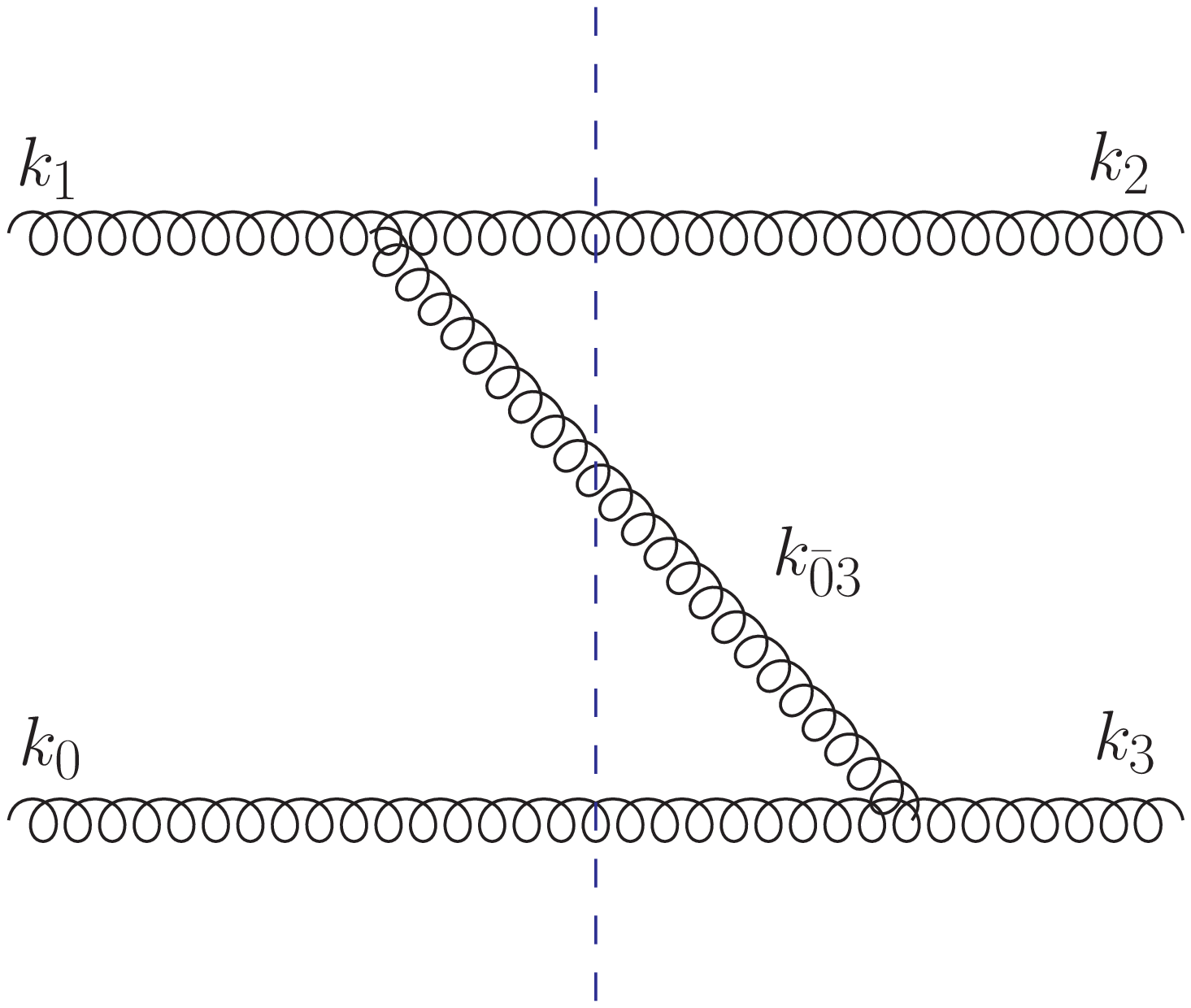}}}\hspace*{0.7cm}
\subfloat[]{\label{fig:2to2c}\includegraphics[width=.25\textwidth]{{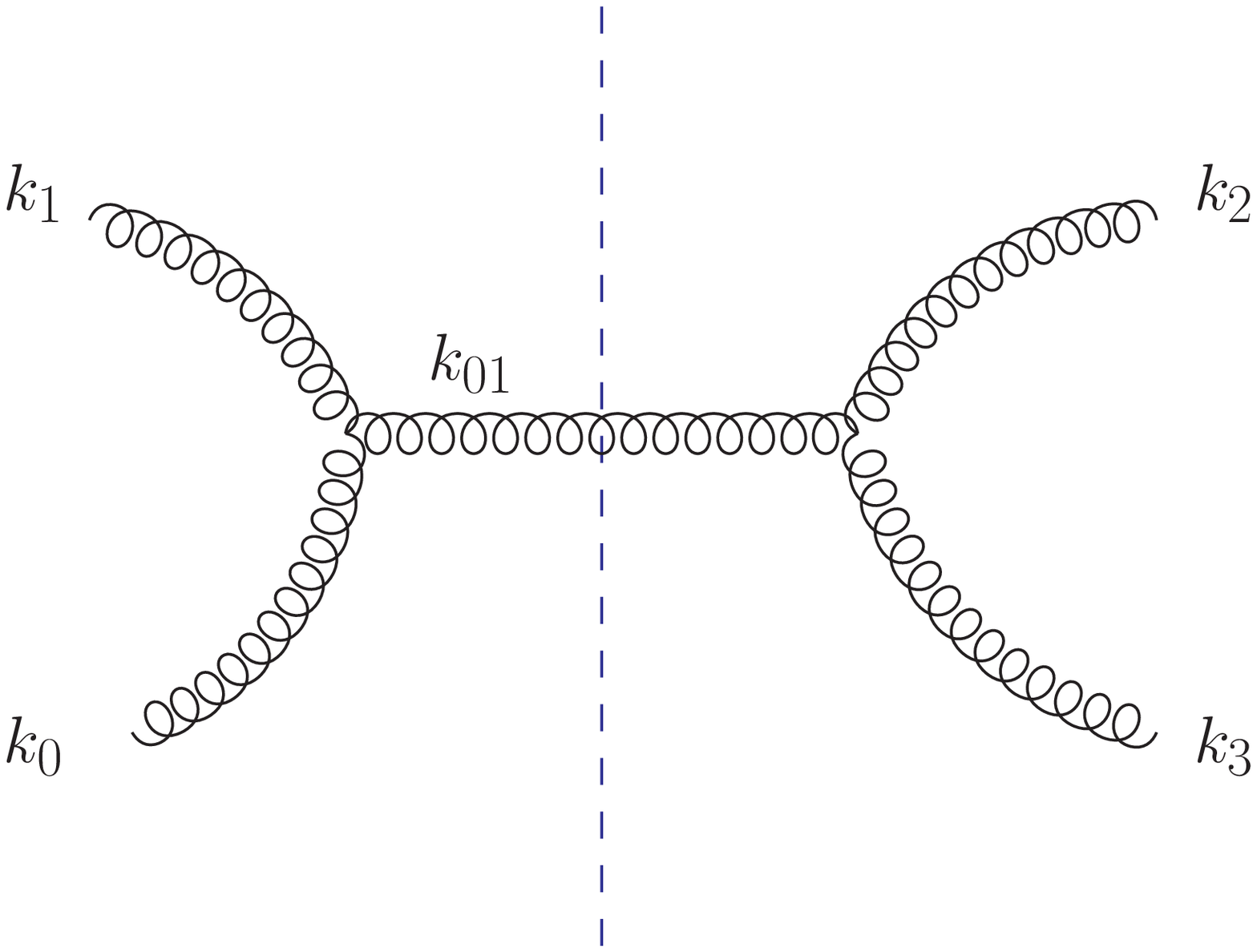}}}\\
\subfloat[]{\label{fig:2to2d}\includegraphics[width=.25\textwidth]{{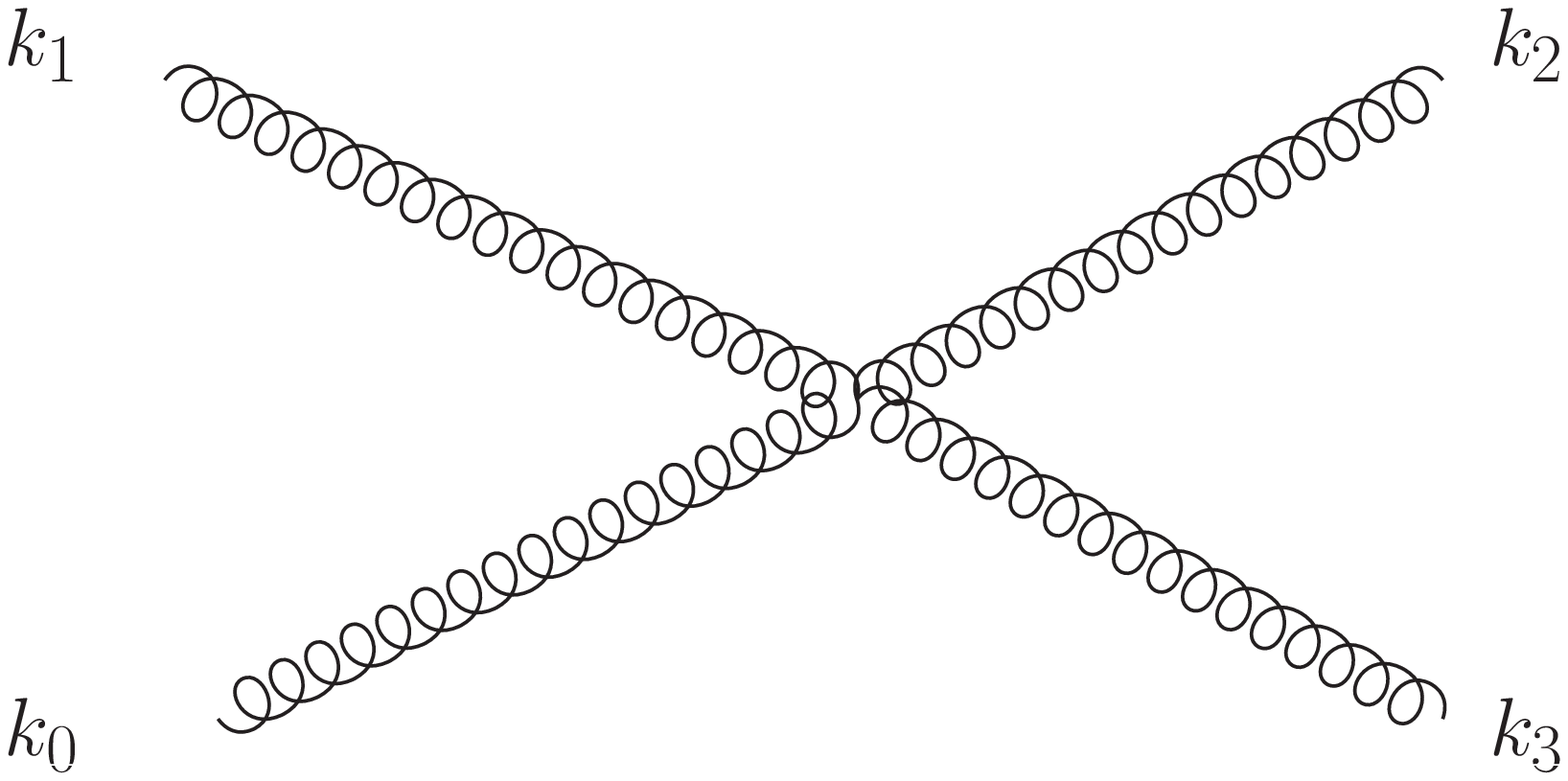}}}\hspace*{0.7cm}
\subfloat[]{\label{fig:2to2e}\includegraphics[width=.25\textwidth]{{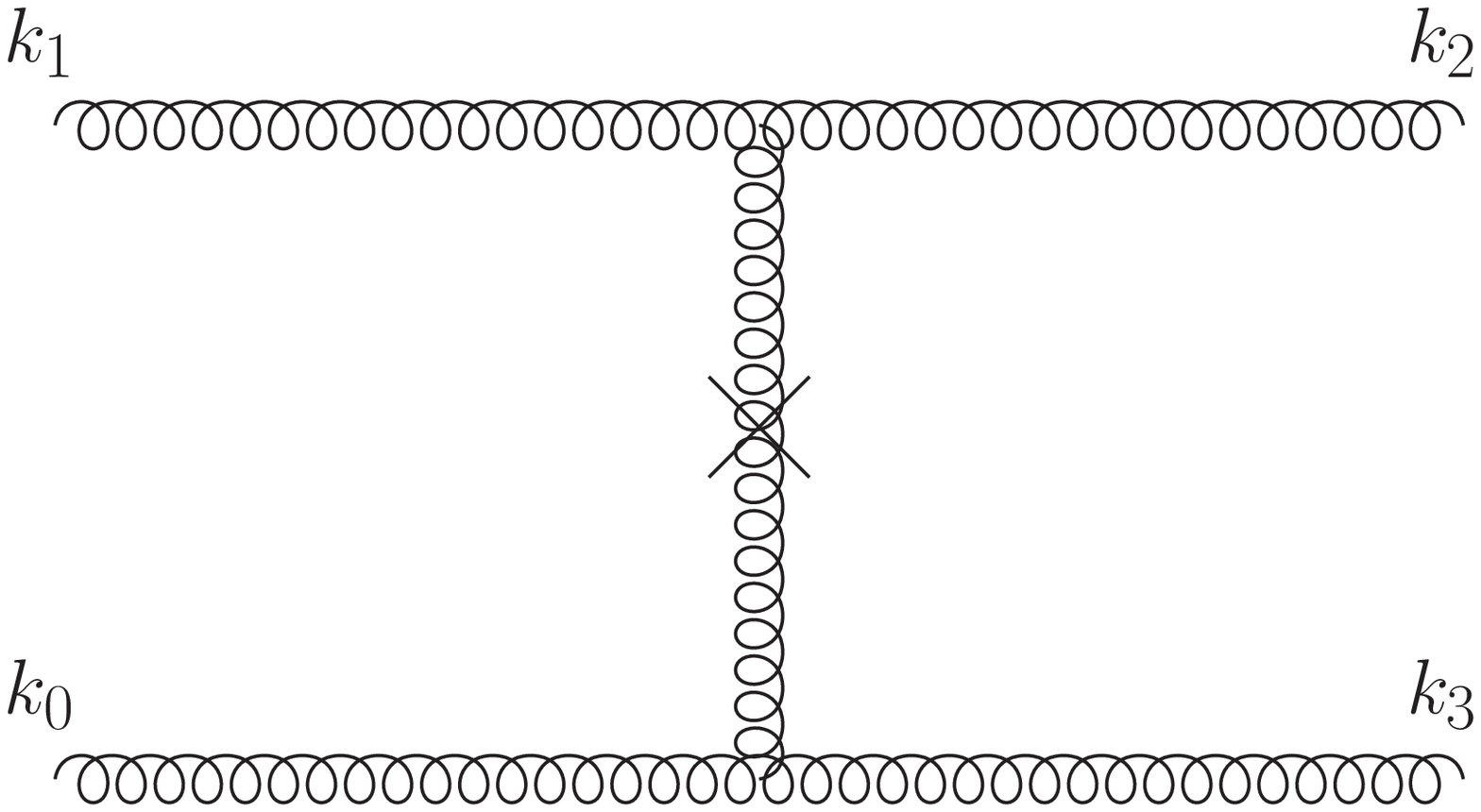}}}
\caption{Basic graphs which contribute to the $2\rightarrow 2$ amplitude in LFPT. Dashed lines indicate the energy denominators for the intermediate states. The cross on the vertical line in graph (e) indicates the instantaneous - Coulomb gluon exchange.  }
\label{fig:2to2graphs}
\end{figure}

To calculate the sub-amplitude for 2 $\to$ 2 gluon scattering $M_{2\to2}$ one needs to compute and add up all the graphs shown in Fig.~\ref{fig:2to2graphs}.   Let us focus on the situation where the helicities for this amplitude are $m(0^+,1^-,2^+,3^+)$, the momenta of the gluons 0 and 1 are incoming and the momenta of the gluons 2 and 3 are outgoing. It is well known that the amplitude for such helicity configuration has to vanish, see \cite{Mangano:1990by} and discussion in the previous section.   Thus it is the simplest example, nevertheless it is instructive to see how that happens on the light-front, using the notation which was given in the previous section.  Graphs $d$ and $e$ give contributions which vanish identically for this helicity configuration. However, the contributions from graphs $a,b,c$ do not vanish separately. The sub-amplitude $G_a$ for the graph in Fig. \ref{fig:2to2a} is given by
\be
 G_a = - \frac{i}{z_{2\bar{1}}}\frac{V_1 V_2}{D_3} \; ,
\label{eq:2to2a1}
\ee
where the vertices and the energy denominator are given by
\be
V_1 = 2 g z_0 \uvec{\epsilon}^{(+)}\cdot \left(\frac{\uvec{k}_{3}}{z_3}-\frac{\uvec{k}_{(2\bar{1})}}{z_{(2\bar{1})}}\right) =  2 g z_0 v^*_{3(2\bar{1})} \; ,
\ee
\be
V_2 = 2g z_{(2\bar{1})} \uvec{\epsilon}^{(+)}\cdot \left(\frac{\uvec{k}_{1}}{z_1}-\frac{\uvec{k}_{2}}{z_1}\right)=-2 g z_{(2\bar{1})}v^*_{21} \; ,
\ee
\be
D_3=-\left(\frac{\uvec{k}_{0}^2}{z_0}-\frac{\uvec{k}_{(2\bar{1})}^2}{z_{(2\bar{1})}}-\frac{\uvec{k}_{3}^2}{z_3}\right) \; ,
\ee
to give
\be
G_a = 4 ig^2  z_0\frac{  -v^*_{3(2\bar{1})}  v^*_{21}}{\frac{\uvec{k}_{0}^2}{z_0}-\frac{\uvec{k}_{(2\bar{1})}^{2}}{z_{(2\bar{1})}}-\frac{\uvec{k}_{3}^2}{z_3}}\; .
\label{eq:2to2a}
\ee
where we used notation $\bar{1}$ to denote the change in the sign of the corresponding transverse and longitudinal momenta i.e. $z_{i\bar{j}} \equiv z_i-z_j$ and $\uvec{k}_{(i\bar{j})} \equiv \uvec{k}_i-\uvec{k}_j$ and $v_{3\;(2\bar{1})}=\frac{\uvec{k}_3}{z_3}-\frac{\uvec{k}_2-\uvec{k}_1}{z_1-z_1}$. Also we have dropped an overall, dimensional factor of $P^+$ from the energy denominator to facilitate the comparison with the previous expressions.
Similarly, the sub-amplitudes $G_b$ and $G_c$ for the graphs in Figs. \ref{fig:2to2b} and \ref{fig:2to2c}, respectively, are given by
\be
G_b =   4ig^2  z_0\frac{ - v^*_{(1\bar{2})3}  v^*_{21}}{\frac{\uvec{k}_{1}^2}{z_1}-\frac{\uvec{k}_{(1\bar{2})}^{2}}{z_{(1\bar{2})}}-\frac{\uvec{k}_{2}^2}{z_2}} \; ,
\label{eq:2to2b}
\ee
\be
G_c = 4i g^2  z_0\frac{  v^*_{\bar{1}(23)}  v^*_{32}}{\frac{\uvec{k}_{0}^2}{z_0}+\frac{\uvec{k}_{1}^{2}}{z_1}-\frac{\uvec{k}_{(23)}^2}{z_{(23)}}} \; .
\label{eq:2to2c}
\ee
We have to take into account that the internal lines need to have the positive fractions of the longitudinal momenta in Figs. \ref{fig:2to2a} and \ref{fig:2to2b}. This means we need to include the corresponding step functions which enforce the ordering of the momenta.
 Hence, $M_{2\to2}$ resulting from these three graphs is given by
\be
M_{2\to2} = \Theta(z_0-z_3) G_a+\Theta(z_3-z_0) G_b+G_c \; .
\ee

It is trivial to show that $G_a$ and $G_b$ are actually the same and can be combined into one expression. Using $\frac{\uvec{k}_{0}^2}{z_0}+\frac{\uvec{k}_{1}^2}{z_1}=\frac{\uvec{k}_{2}^2}{z_2}+\frac{\uvec{k}_{3}^2}{z_3}$ we can write the denominator in \eqref{eq:2to2b} as $-\frac{\uvec{k}_{0}^2}{z_0}-\frac{\uvec{k}_{(1\bar{2})}^{2}}{z_{(1\bar{2})}}+\frac{\uvec{k}_{3}^2}{z_3}$.  The following identities also hold
\be
v_{(1\bar{2})3}=-v_{3(2\bar{1})}, \; \; z_{(1\bar{2})}=-z_{(2\bar{1})} \; .
\ee
Using them in \eqref{eq:2to2a} and \eqref{eq:2to2b} we realize that the expressions for $G_a$  and $G_b$ are the same which  gives
\be
M_{2\to2} =  G_a+G_c \;,
\label{eq:2to2}
\ee
where the theta functions are absent (only global momentum conservation conditions on external longitudinal momenta are present but they are implicit).
The fact that the graphs $a$ and $b$ give the same contributions, can be shown to hold
for more complicated graphs. This is discussed in detail in Appendix A and it will be used later to prove the relationships between wave functions and scattering amplitudes.

 Having reduced the expression for the sum of both graphs to \eqref{eq:2to2} we proceed to show that    amplitude $ M_{2\rightarrow 2}$ in (\ref{eq:2to2}) has identical structure to the form of the gluon wave function $\Psi_3$ in Eq.~\eqref{eq:wavefunction3} (modulo overall factors). To demonstrate this  one can  use the relation $D_3=\xi_{3(2\bar{1})}|\uvec{v}_{3(2\bar{1})}|^2$ to obtain
\be
G_a = -2 ig^2 z_0 \frac{v_{21}^*}{\xi_{3\;(2\bar{1})}v_{3\;(2\bar{1})}} \; ,
\ee 

Similarly, we can rewrite the energy denominator of graph (c)  $$\frac{\uvec{k}_{0}^2}{z_0}+\frac{\uvec{k}_{1}^{2}}{z_1}-\frac{\uvec{k}_{(23)}^2}{z_{(23)}}=-\xi_{(32)\bar{1}} |\uvec{v}_{(32)\bar{1}}|^2\; ,$$to get
\be
G_c =  - 2 ig^2z_0 \frac{v_{32}^*}{\xi_{(32)\bar{1}}v_{(32)\bar{1}}} \; . 
\ee
Putting both expressions together we obtain, 
\begin{align}
M_{2\to2}&= -2i g^2 z_0 \left( \frac{v_{21}^*}{\xi_{3\;(2\bar{1})}v_{3\;(2\bar{1})}}+\frac{v_{32}^*}{\xi_{(32)\bar{1}}v_{(32){\bar{1}}}}\right) \\
&=-2i g^2 z_0\left(\frac{v_{21}^*\xi_{(32)\bar{1}}v_{(32)\bar{1}}+v_{32}^*\xi_{3\;(2\bar{1})}v_{3\;(2\bar{1})}}{\xi_{3\;(2\bar{1})}\xi_{(32)\bar{1}}v_{3\;(2\bar{1})}v_{(32)\bar{1}}}\right) \; .
\label{eq:M2to2num}
\end{align}
We notice that the structure of $M_{2\to2}$ now has  exactly  the same structure as the wave function for  $\Psi_{3}$ in \eqref{eq:wavefunction3}, when replacing $\bar{1}\rightarrow 1$.  
The only difference is the overall normalization factor due to slightly different convention used in \cite{ms} to derive \eqref{eq:wavefunction3}.
Most importantly, just as in \eqref{eq:wavefunction3}, the numerator in \eqref{eq:M2to2num} is proportional to the difference of the energies from the initial to the last state in the problem.  Since in the case considered (i.e. on-shell amplitude) the last state is on-shell (final) state, the numerator, and thus $M_{2\to2}$, will be zero.
Therefore we see that the relatively compact form of the gluon wave functions derived in \cite{ms} and the recursion relations for these wave functions are closely related to the properties of the amplitudes. In this particular case the simplicity of the wave functions are related to  the fact that the amplitudes with a single helicity different from the rest are vanishing. The result derived for lowest order scattering can be generalized to arbitrary number of the final state particles (at the tree-level). 
  It is important to note though that the wave functions derived in Sec.~\ref{sec:wvfr} themselves are non-zero due to the off-shellness of the last state.

\subsection{General relations between the wave functions and the amplitudes}

Given the result of the explicit example shown in the previous section we proceed to investigate the relation between the amplitude for a general number of final state gluons $M_{2\to n}$ and the wave function $\Psi_{n+1}$ which were derived previously, see Ref.~\cite{ms} and Sec.~\ref{sec:wvfr}. That such relation should exist is quite intuitive as the two objects correspond to the graphs which are topologically equivalent with only the one external leg changing the momentum from the final to initial state.  In the case of the LFPT it is however not an entirely obvious relation as the energy denominators, which entangle the momenta of the particles in a given graph,  in principle change via such operation and, moreover, some graphs are vanishing. For example,
for the case of the $\Psi_3$ we have only two graphs depicted in Fig.~\ref{fig:wf3} which contain the 3-gluon vertices and give non-trivial contributions, whereas for the $2\to2$ amplitude there are three graphs (with 3-gluon vertices) whose contributions are non-vanishing, those shown in Fig.~\ref{fig:2to2a}, \ref{fig:2to2b}, \ref{fig:2to2c}.
Therefore by changing the momentum line from outgoing to incoming in Fig.~\ref{fig:wf3}, one needs to perform additional summation over the time-ordering of some of the vertices, so that in our example the contribution from graph (a) in Fig.~\ref{fig:wf3} has to  give two separate contributions in Fig.~\ref{fig:2to2a} and \ref{fig:2to2b}. We shall show how to identify and sum over the time-ordering for the case of $2\to3$ and then we perform a proof for the general case and show that
the amplitude from the $M_{2\to n}$ graphs can be obtained through analytical continuation from  the  light-front wave functions $\Psi_{n+1}$ which contain in general a smaller number of graphs. 

We will restrict ourselves to the case of only 3-gluon vertices which will limit the possible helicity states. However, the methods developed are much more general and we will utilize them later to find amplitudes with more complicated helicity states.

Let us first consider  the case of 2 to 3 gluon scattering.  We want to find the $M_{2\to3}$ graphs from the graphs of the gluon wave function which describes 1 to 4 gluon transition in the $+\rightarrow ++\dots+$ configuration. This  gluon wave function contains $(4-1)!=6$ graphs. 
\begin{figure}[h]
\centering
\includegraphics[width=.8\textwidth]{{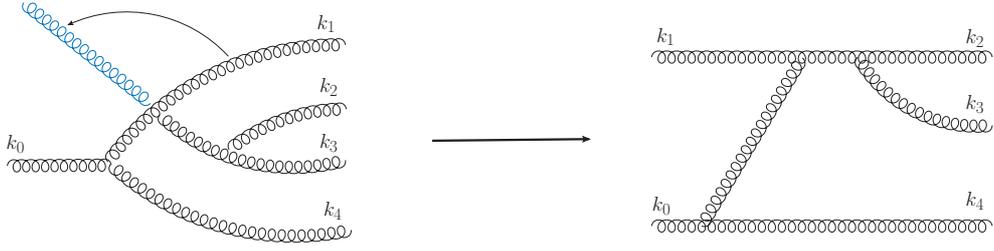}}
\caption{By changing the outgoing to incoming momentum one can recover the graphs of the amplitudes from the wave functions.}
\label{fig:conversion}
\end{figure}

\begin{figure}
\centering
\subfloat[Subgroup $G_{1,1}$]{\includegraphics[height=.1\textheight]{{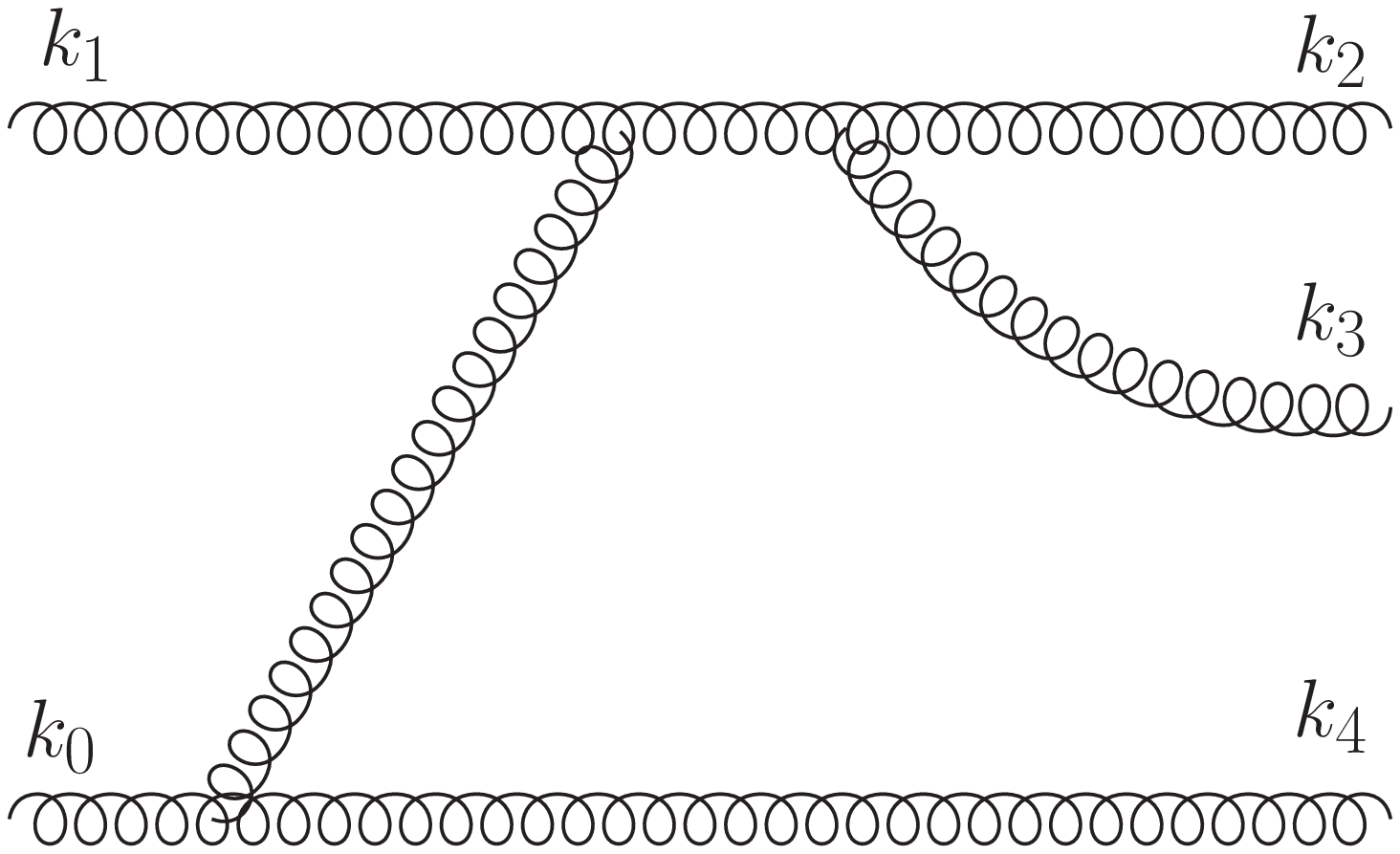}}}
\qquad\qquad\qquad\qquad\subfloat[Subgroup $G_{1,2}$]{\includegraphics[height=.1\textheight]{{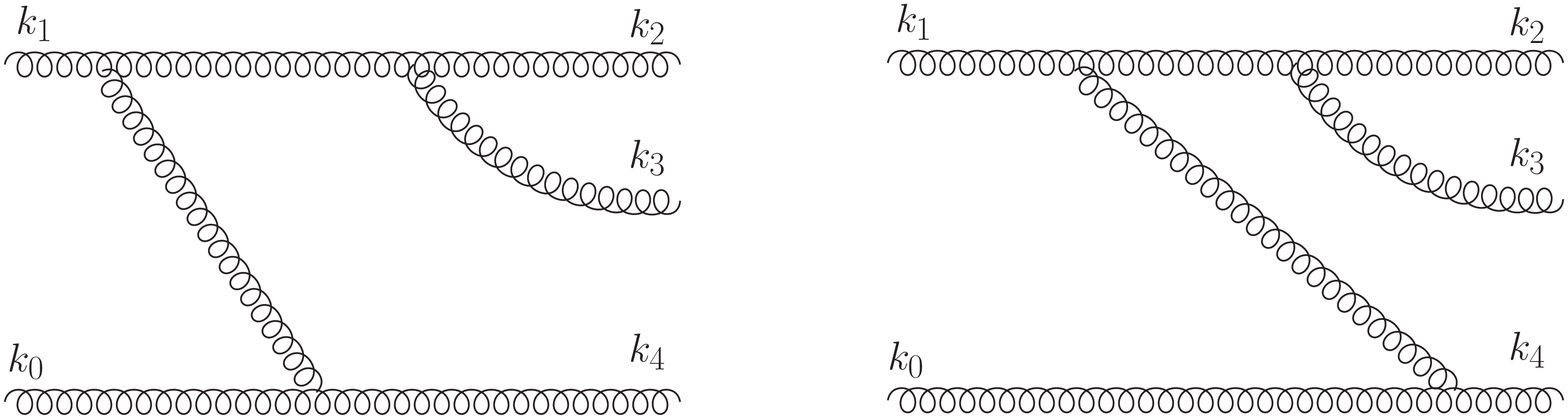}}}
\caption{A group of topologically equivalent graphs for $2\to3$ amplitude obtained from the continuation of the $k_1$ momentum, as shown in Fig. \ref{fig:conversion}.}
\label{fig:some2to3graphs}
\end{figure}

\begin{figure}
\centering
\subfloat[Subgroup $G_{2,1}$]{\includegraphics[height=.1\textheight]{{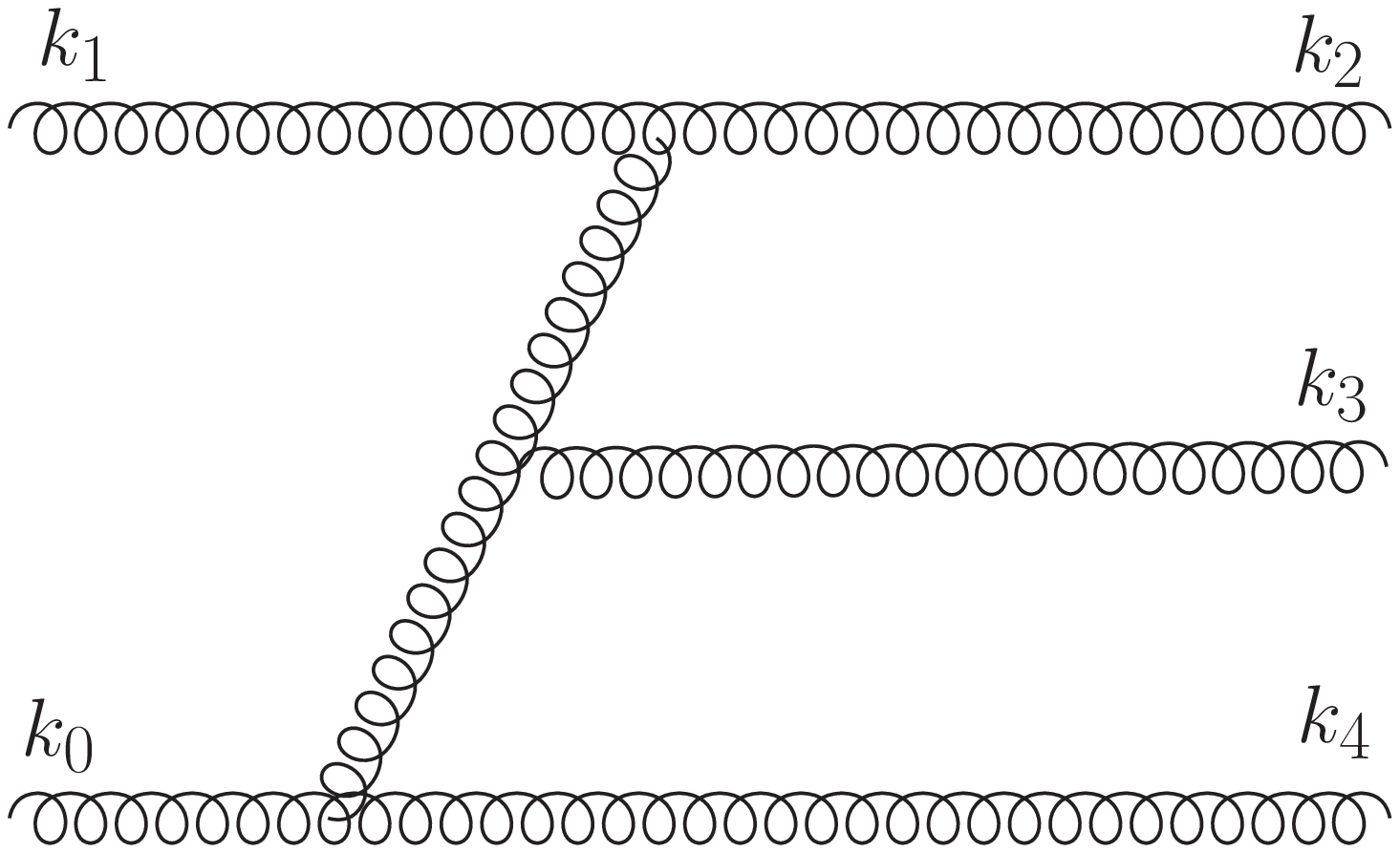}}}
\qquad\qquad\qquad\qquad\subfloat[Subgroup $G_{2,2}$]{\includegraphics[height=.1\textheight]{{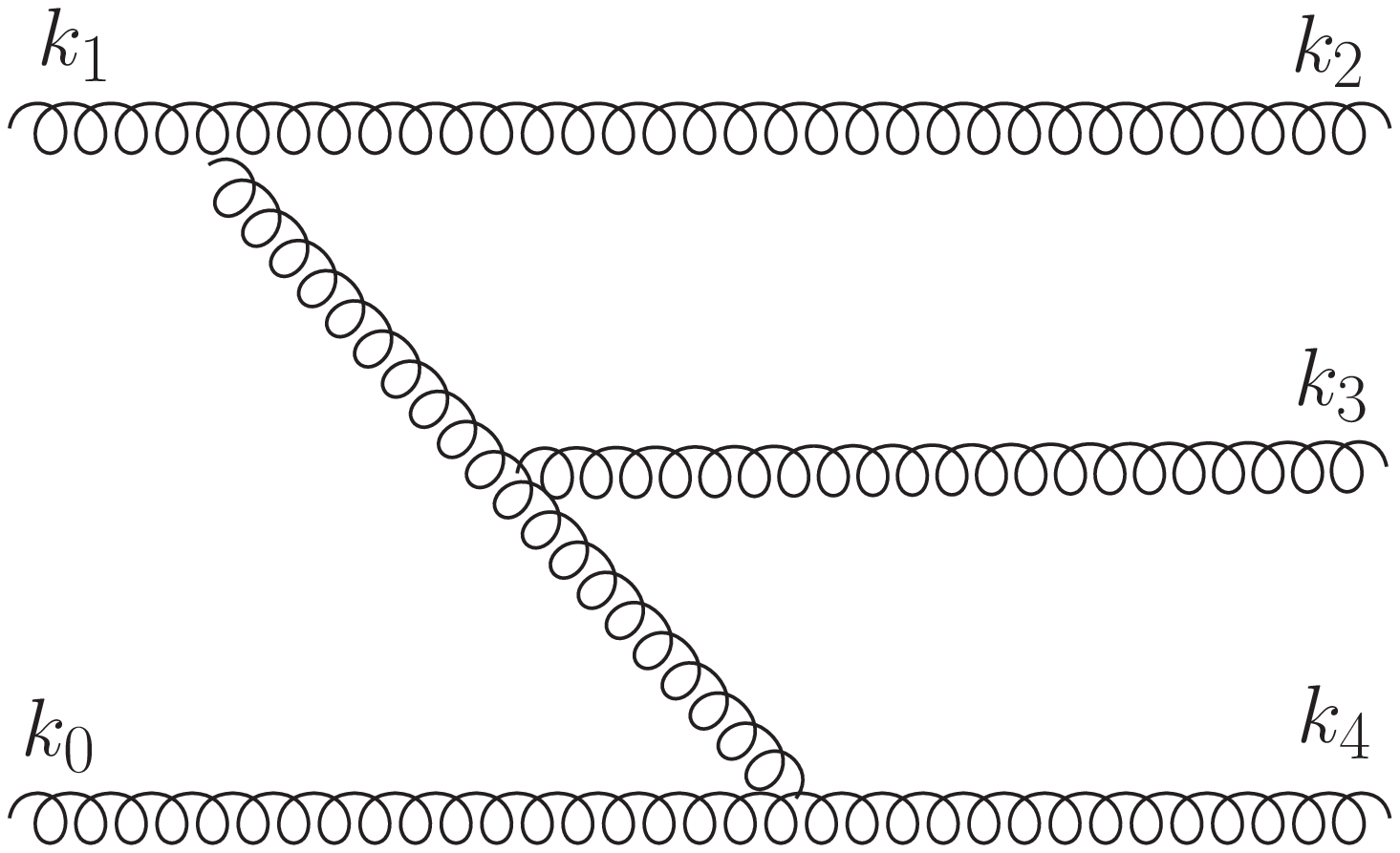}}}
\\\subfloat[Subgroup $G_{2,3}$]{
\begin{minipage}[c]{0.5\linewidth}
\centering
\includegraphics[height=.1\textheight]{{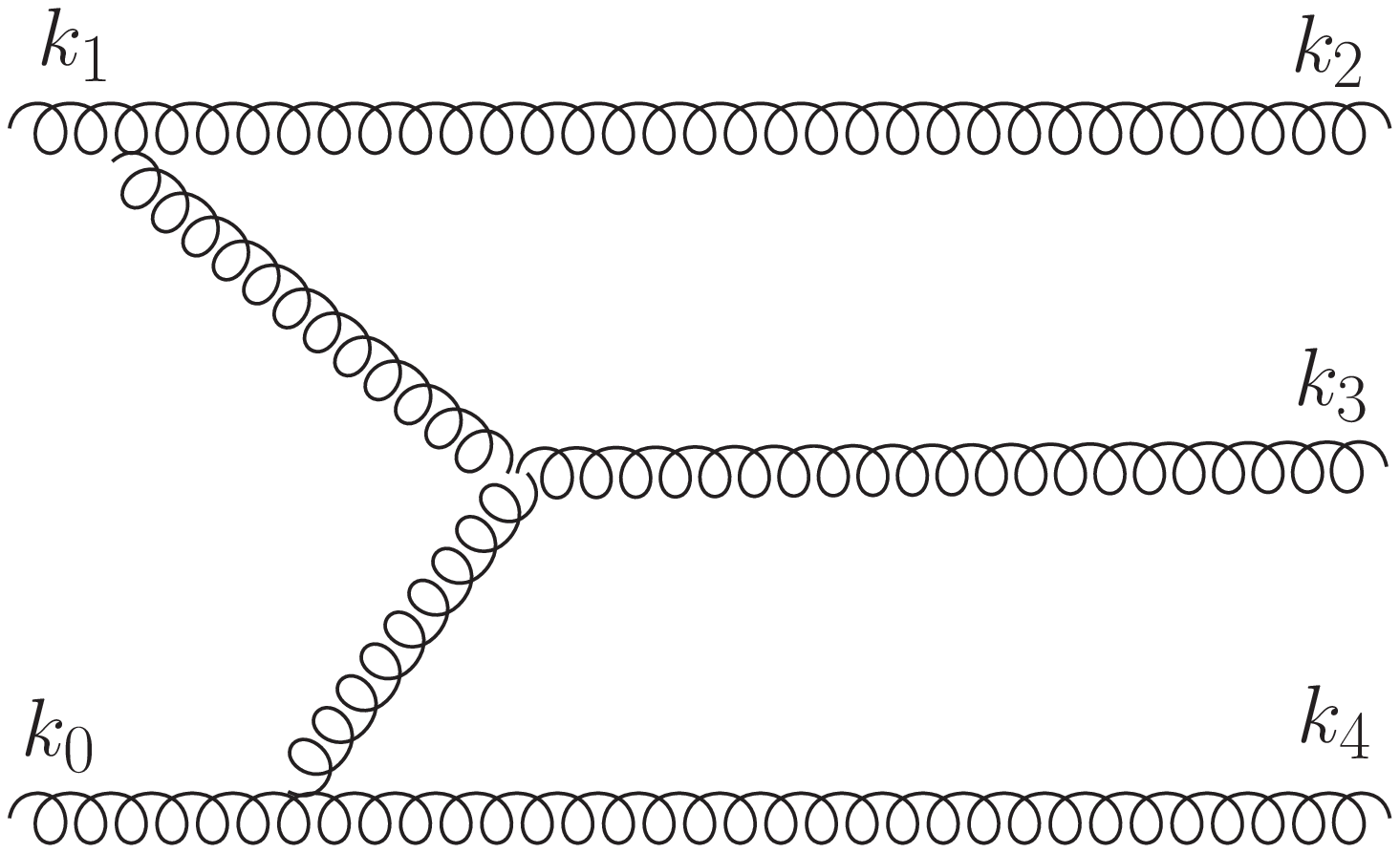}}
\end{minipage}
\negthickspace\negthickspace
\begin{minipage}[c]{0.5\linewidth}
\centering
\includegraphics[height=.1\textheight]{{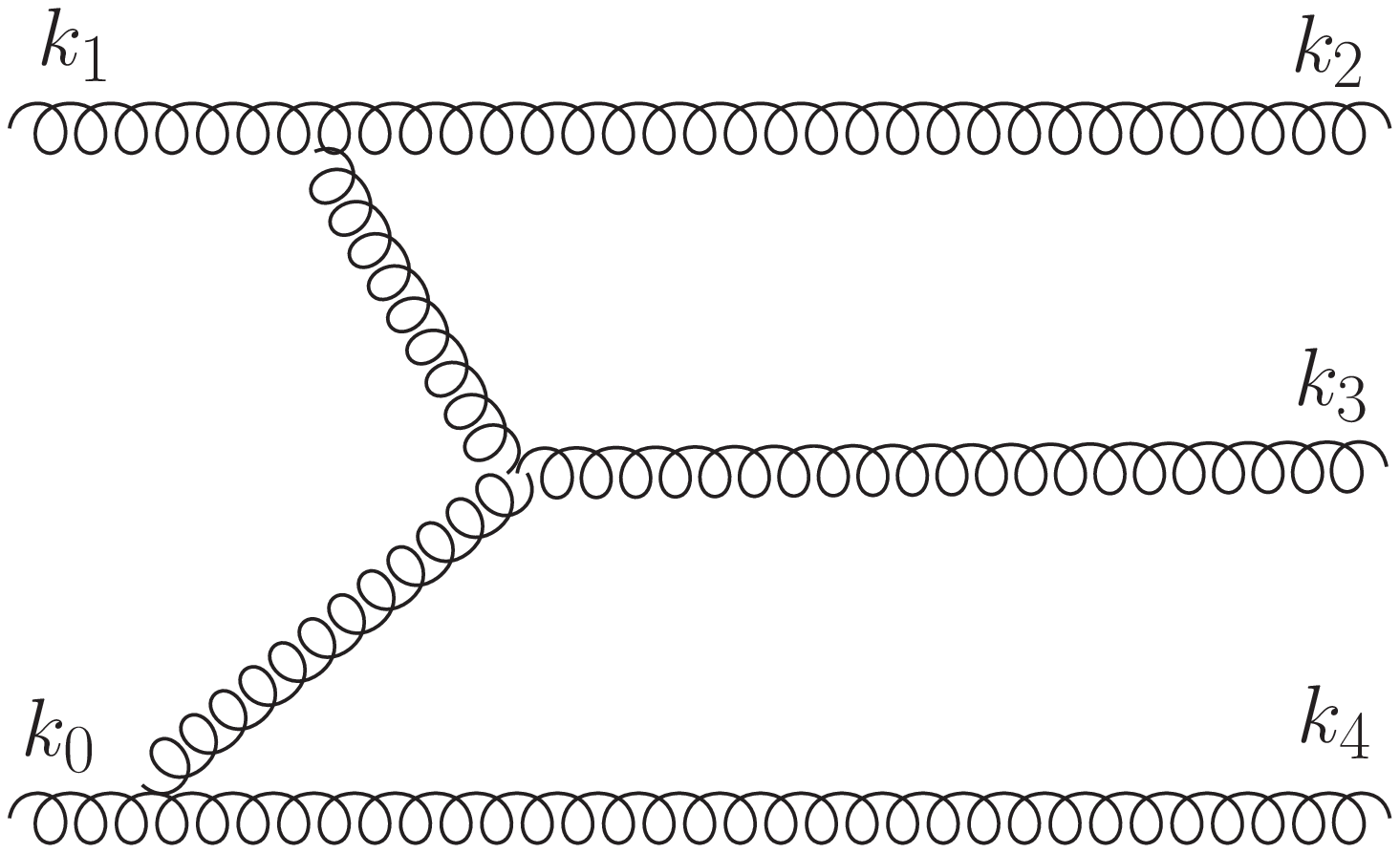}}
\end{minipage}}
\caption{A group of topologically equivalent graphs for $2\to3$ amplitude with different time-orderings.}
\label{fig:group1}
\end{figure}

\begin{figure}
\centering
\subfloat[Subgroup $G_{3,1}$]{
\begin{minipage}[c]{0.5\linewidth}
\centering
\includegraphics[height=.1\textheight]{{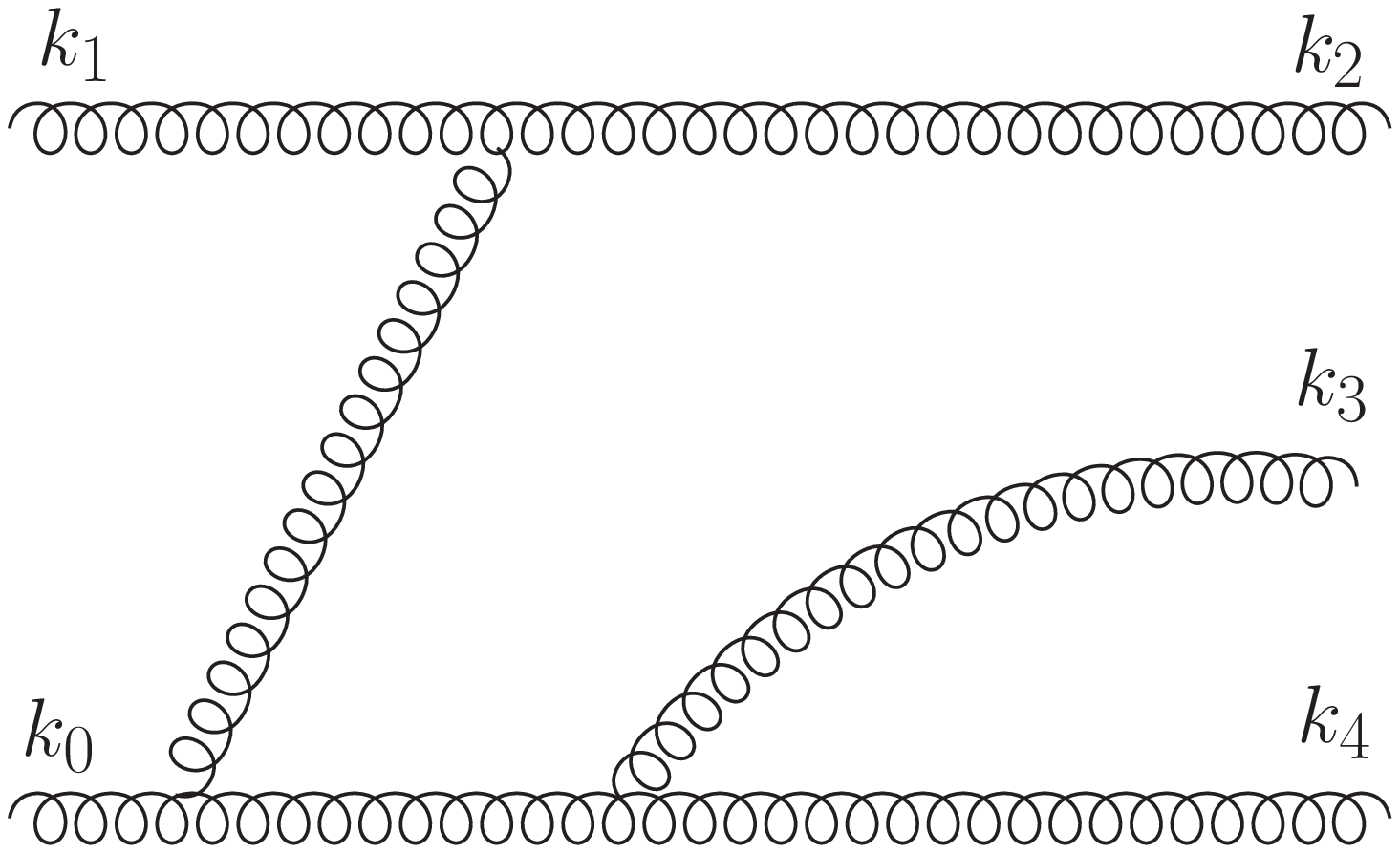}}
\end{minipage}
\begin{minipage}[c]{0.5\linewidth}
\centering
\includegraphics[height=.1\textheight]{{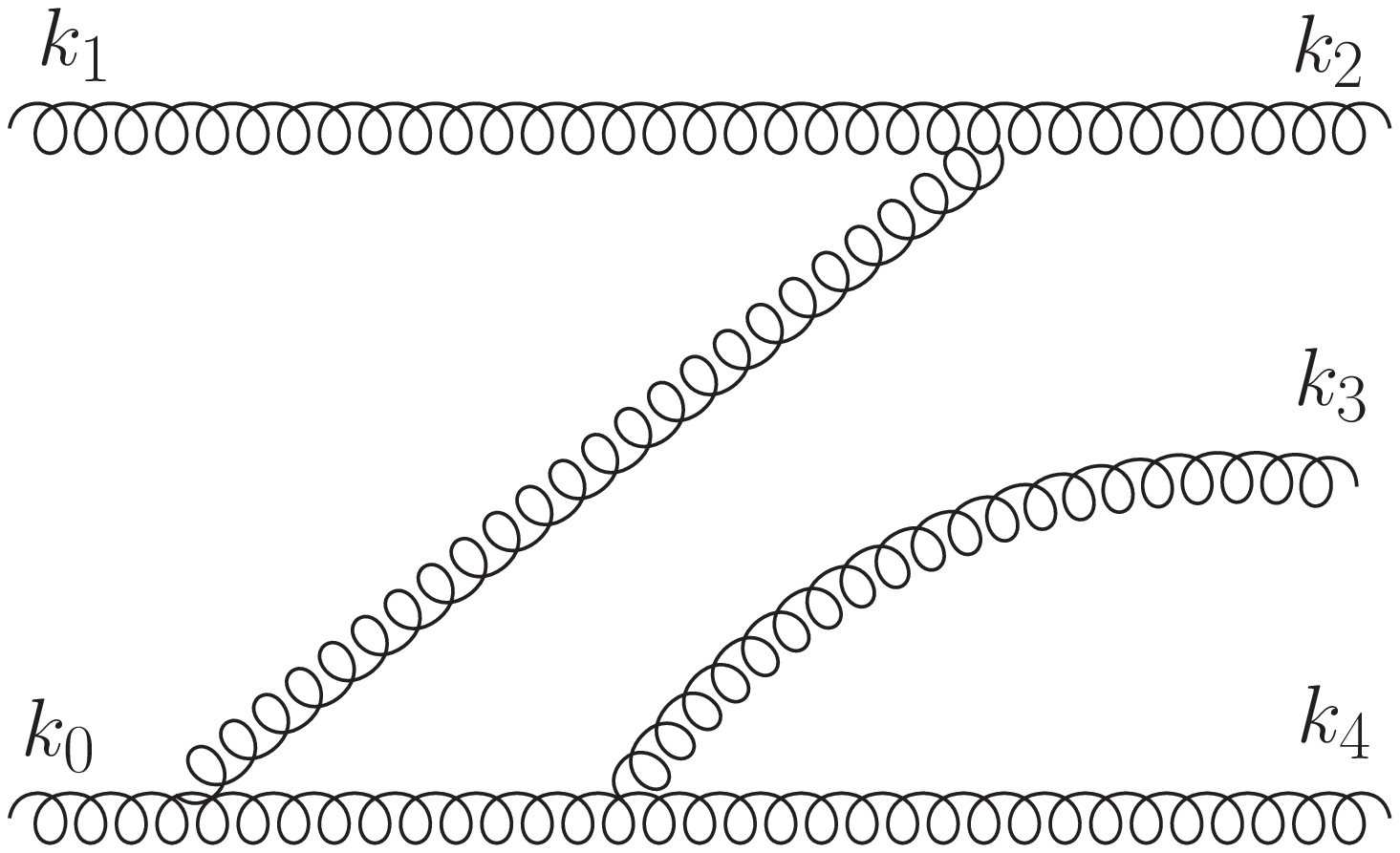}}
\end{minipage}}
\\\subfloat[Subgroup $G_{3,2}$]{\includegraphics[height=.1\textheight]{{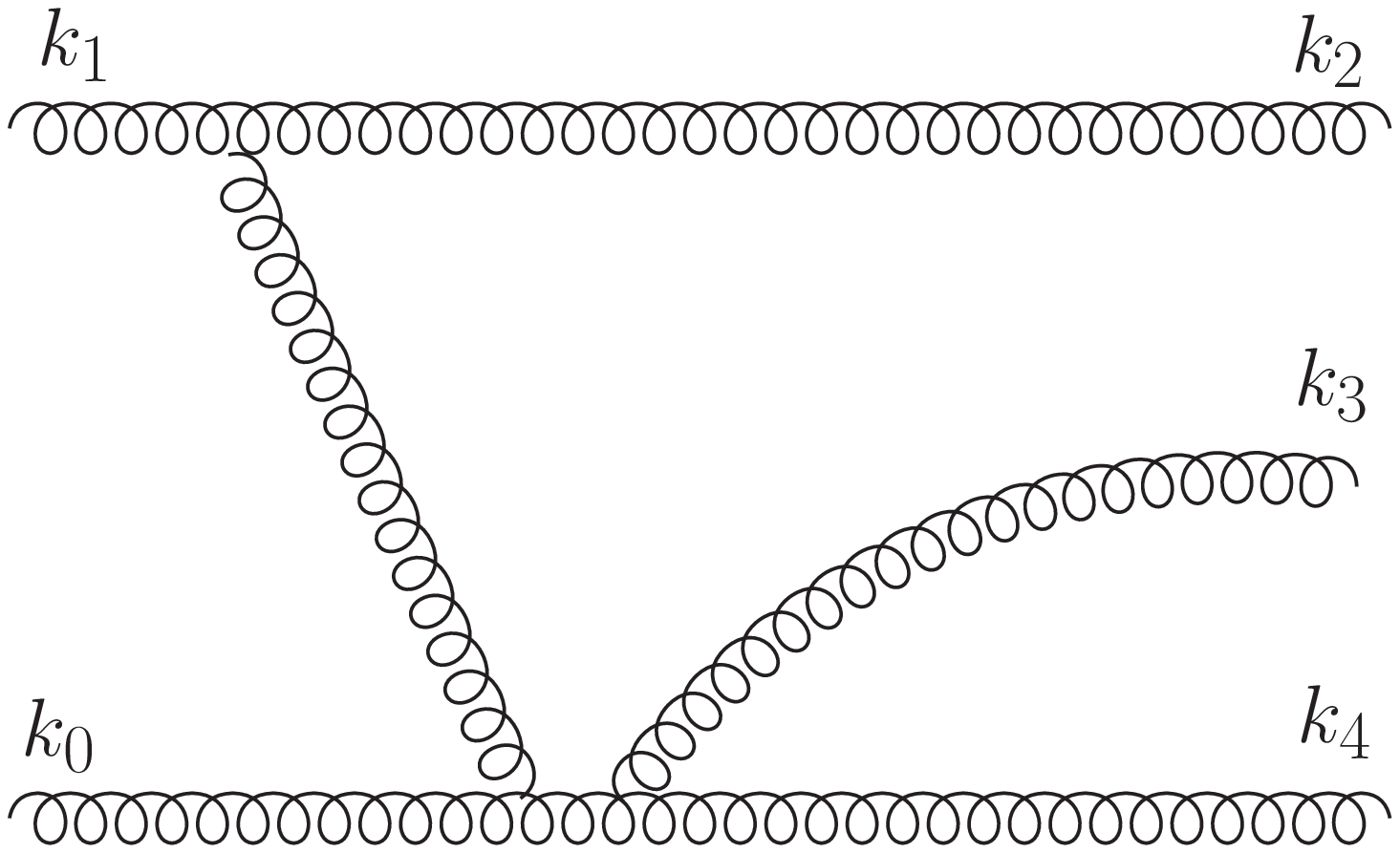}}}
\caption{A group of topologically equivalent graphs for $2\to3$ amplitude with different time-orderings.}
\label{fig:group2}
\end{figure}

\begin{figure}
\begin{minipage}[c]{0.5\linewidth}
\centering
\includegraphics[height=.1\textheight]{{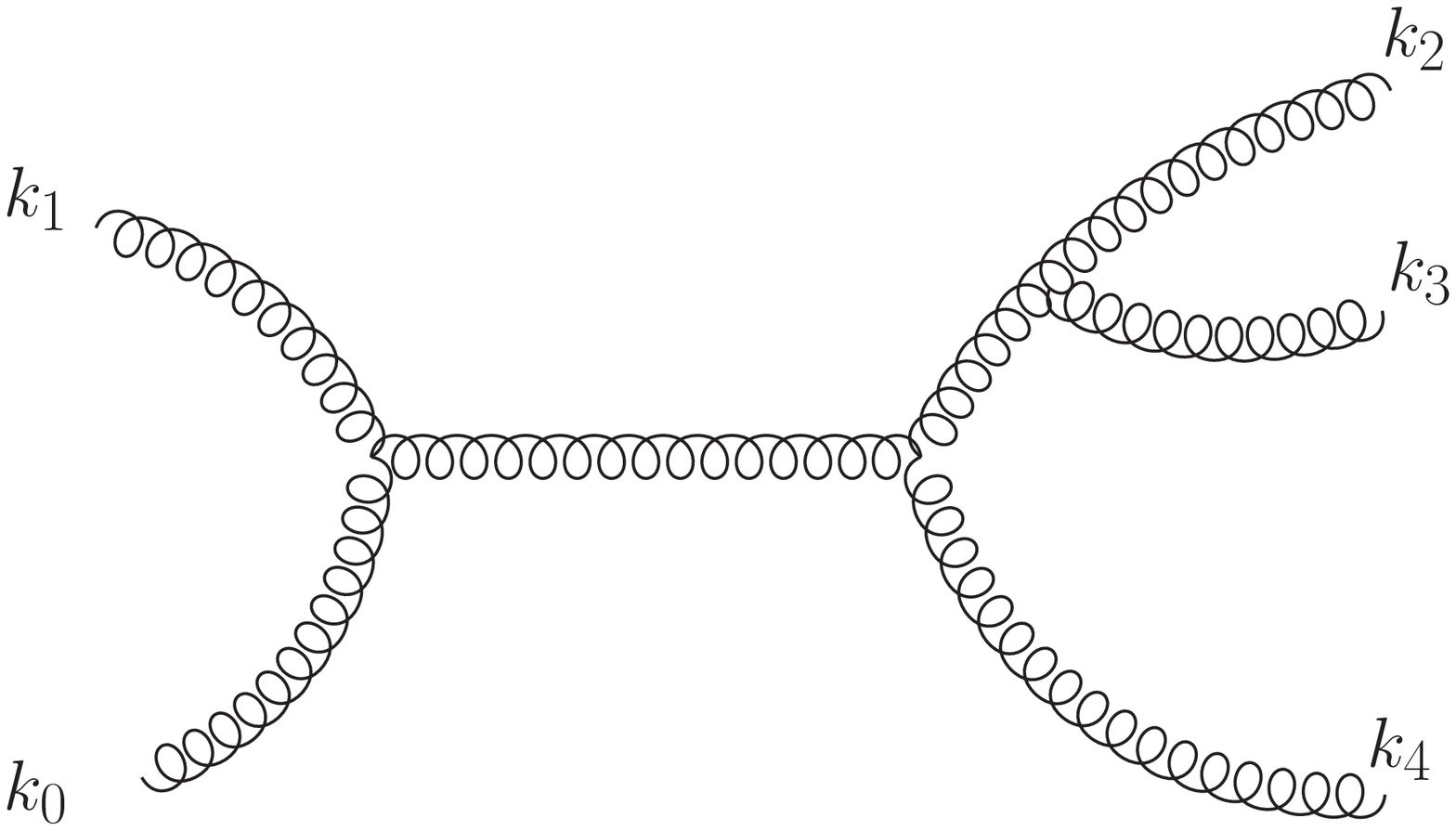}}
\label{fig:group4a}
\end{minipage}
\begin{minipage}[c]{0.5\linewidth}
\centering
\includegraphics[height=.1\textheight]{{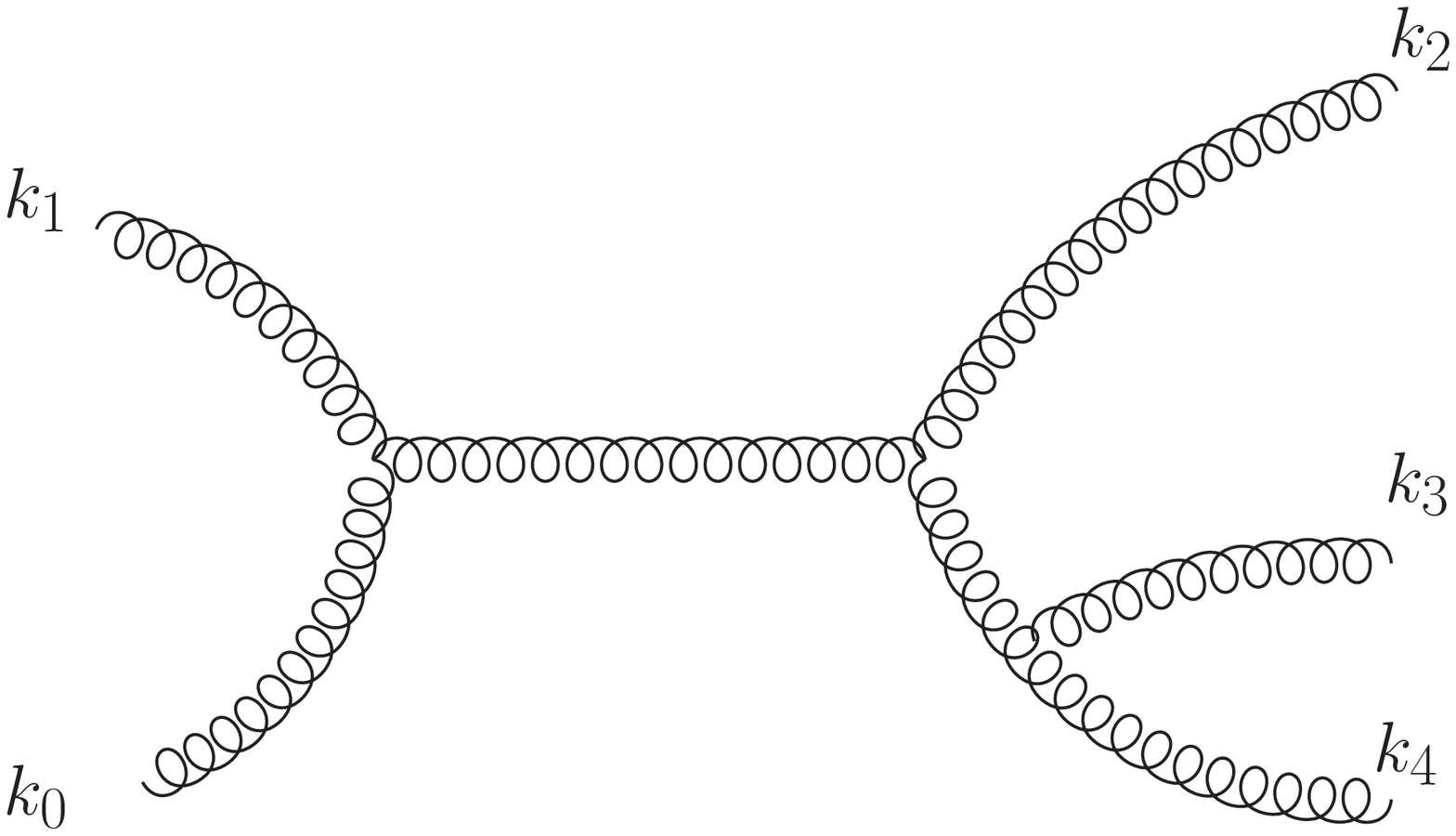}}
\label{fig:group4b}
\end{minipage}
\caption{$s$ - channel graph contributions to $2\to3$ scattering.  Groups $G_5$ and $G_6$ are shown on the left and right respectively.  Both consist of only one graph.}
\label{fig:group4}
\end{figure}

 The way to obtain graphs for $2\to 3$ scattering is by changing the $k_1$ line from outgoing into incoming from the gluon wave function, as shown in Fig. \ref{fig:conversion}.  Naively, one recovers only one graph from the wave functions, which is shown in this figure. However, since our calculation is being done in the LFPT, we need to find all possible time-orderings.  Thus, from the sole graph depicted on the left hand side of  Fig.~\ref{fig:conversion} we obtain three graphs  of the same  topology,  which are depicted in Fig.~\ref{fig:some2to3graphs}. This continuation and summation needs to the performed   with all the graphs in the $\Psi_4$ wave function to arrive at all the corresponding graphs needed to calculate $M_{2\to3}$, depicted in Figs.~\ref{fig:some2to3graphs}-\ref{fig:group4}.  Here we have excluded all the vacuum diagrams since they vanish on the light-front.


Figs. \ref{fig:some2to3graphs}-\ref{fig:group4} group together all the graphs with the same topology.  These groups are in turn composed of subgroups where the directions of the lines are taken into account.  We know that for a graph to have a non-zero value all its lines must have a positive longitudinal momentum. Thus, different kinematical conditions which are satisfied by the longitudinal momenta of the external lines apply for the different subgroups inside a group.  
To make this statement more clear we can introduce  the following classification for the  graphs. Let us call $G_i$ the group of graphs which have the same topology, with $G_2$ being the group depicted in Fig.~\ref{fig:group1}. We denote by $G_{i,j}$ the subgroup of graphs in group $G_i$ which are distinguished by different kinematical conditions on the external momenta. In our example $G_{2,j}$ denote the subgroups listed in plots a-c in Fig.~\ref{fig:group1}. 

Let
\be
s_{2,j}\equiv \text{ sum of graphs in subgroup }G_{2,j} \; ,
\ee
\be
S_2\equiv \text{ sum of graphs in group }G_2 \; .
\ee

We can  then write 
\be
S_2 = s_{2,1} \Theta(z_0-z_4)\, \Theta(z_2-z_1) + s_{2,2} \Theta(z_4-z_0)\,\Theta(z_1-z_2) + s_{2,3} \Theta(z_0-z_4) \, \Theta(z_1-z_2) \; .
\label{eq:s4}
\ee
However, it can be demonstrated  that contributions from all the graphs are actually the same, i.e. $s_2 \equiv s_{i,1}=s_{2,2}=s_{2,3}$ (this is discussed  in Appendix for a general case). Thus,
\be
S_2 = s_2 \; .
\ee
This sum contains all physical arrangements of the kinematical conditions on the external lines, so we can remove the $\Theta$ functions.

We should  note that some groups consist of only a single subgroup.  These will be composed of only $s$-channel graphs and the internal lines are guaranteed to have a positive longitudinal momentum regardless of the external lines. These are graphs shown in Fig.~\ref{fig:group4}. In every group (including the ones with a single subgroup), all of the $m$ graphs in one of the subgroups can be obtained from $m$ distinctive graphs belonging to the wave function $\Psi_4$.  Furthermore, if the graphs in one group can be obtained from a certain set of graphs belonging to $\Psi_4$, then the graphs in another group have to be obtained from a completely different set of graphs. 
Since the value of all subgroups inside a group are the same, in order to calculate $M_{2\to n}$ all we need are the $n!$ graphs that can be obtained by simply changing the direction of the $k_1$ line in each of the $n!$ graphs in  the  1 to $n+1$ gluon wave function $\Psi_{n+1}$.  
We demonstrate this step for arbitrary number of gluons below.
We shall show that the expression for the value of a graph $G$ is equal to the negative value of the expression of the 1 to $n+1$ graph $H$ from which $G$ was obtained.  In general we know, Ref.~\cite{Bjorken:1970ah}, that a graph will be equal to (modulo overall factors)
\be
\frac{\cal{V}}{\cal{Z}\cal{D}} \; ,
\ee
with
\be
{\cal{V}}\equiv\prod_{\text{all vertices}}v_k \;, 
\ee
\be
{\cal{Z}}\equiv\prod_{\text{internal lines}}z_i \; ,
\ee
\be
{\cal{D}}\equiv\prod_{\text{intermediate states}}D_i \; ,
\ee
Thus, we need to compare $\cal{Z}$, $\cal{V}$ and $\cal D$ for $G$ and $H$.

It is important to set up the physical conditions for each of the cases.  These conditions are listed in Table \ref{tab:physconds}.  Note that we have included the energy condition, which we have arrived at by setting the initial and final states on-shell.

\begin{table}
\centering
\scalebox{1.2}{
\begin{tabular}{c c}
2 to $n$ & 1 to $n+1$\\ \midrule
$\uvec{k}_{0}+\uvec{k}_{1} = \sum_{i=2}^{n+1} \uvec{k}_{i}$&$\uvec{k}_0= \sum_{i=1}^{n+1} \uvec{k}_{i}$\\
\noalign{\smallskip}
$z_0+z_1 = \sum_{i=2}^{n+1} z_i$&$z_0= \sum_{i=1}^{n+1} z_i$\\
\noalign{\smallskip}
$\frac{\uvec{k}_{0}^2}{z_0}+\frac{\uvec{k}_{1}^2}{z_1} = \sum_{i=2}^{n+1} \frac{\uvec{k}_{i}^2}{z_i}$&$\frac{\uvec{k}_{0}^2}{z_0} = \sum_{i=1}^{n+1} \frac{\uvec{k}_{i}^2}{z_i}$\\
\noalign{\smallskip}
\bottomrule
\end{tabular}
}
\caption{Physical conditions for gluon scattering for the case of the $2\to n$ and $1\to n+1$ amplitudes.}
\label{tab:physconds}
\end{table}

We shall introduce two variables $k_A$ and $z_A$.  These will be defined differently for the two cases we are working on. For the $2 \to n$ case
$
\uvec{k}_{A}\equiv-\uvec{k}_{1}, \;
z_A\equiv-z_1$
whereas for $1 \to n+1$ case 
$
\uvec{k}_{A}\equiv \uvec{k}_{1}, \;
z_A\equiv z_1.
$
We can now see that these definitions allow us to use the same mathematical expression for both cases.  To be precise, for both 2 to $n$ and 1 to $n+1$ the physical conditions are
\begin{equation*}
\uvec{k}_{0}=\uvec{k}_{A} + \sum_{i=2}^{n+1} \uvec{k}_{i}
\end{equation*}
\begin{equation*}
z_0 =z_A+ \sum_{i=2}^{n+1} z_i
\end{equation*}
\begin{equation}
\frac{\uvec{k}_{0}^2}{z_0}= \frac{\uvec{k}_{A}^2}{z_A} +\sum_{i=2}^{n+1} \frac{\uvec{k}_{i}^2}{z_i}.
\label{eq:physconds}
\end{equation}

Now we can proceed to compare factors $\cal Z$.  Let's take as an example the graphs in Fig. \ref{fig:conversion}.  Looking at these it is clear that, using $k_A$ and $z_A$, we can write, for both cases,
\be
\uvec{k}_{l} = \uvec{k}_{0}-\uvec{k}_{4}=\uvec{k}_{A}+\uvec{k}_{2}+\uvec{k}_{3}, \quad z_l = z_0-z_4=z_A+z_2+z_3,
\ee
\be
\uvec{k}_{q} = \uvec{k}_{0}-\uvec{k}_{4}-\uvec{k}_{A}=\uvec{k}_{2}+\uvec{k}_{3}, \quad z_q = z_0-z_4-z_A=z_2+z_3.
\ee
Once again we see that we can use the same expression for both cases.  This is true for any internal line one considers.  Thus, $\cal Z$ does not change from graphs $2\to n$ and $1\to n+1$.

\begin{figure}[h]
\centering
\subfloat[]{\label{fig:1ton+1denom}\includegraphics[width=.5\textwidth]{{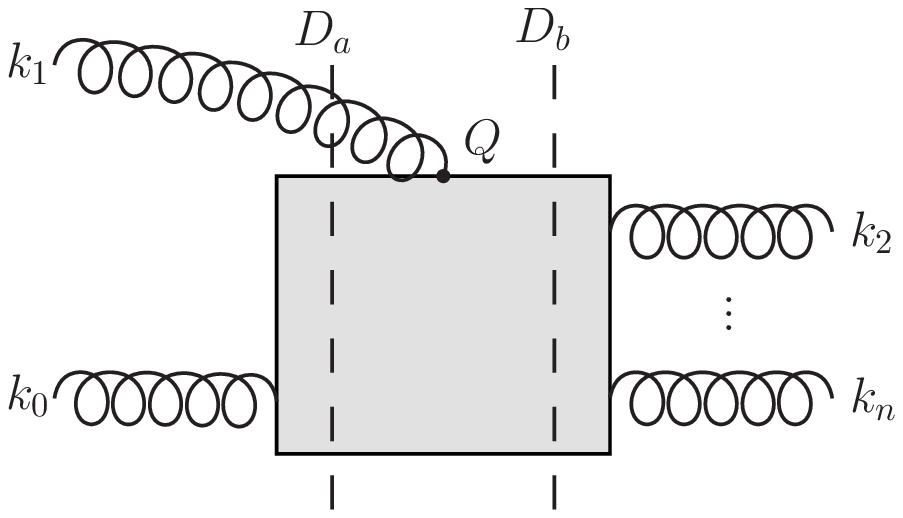}}}
\;\;\;\subfloat[]{\label{fig:2tondenom}\includegraphics[width=.5\textwidth]{{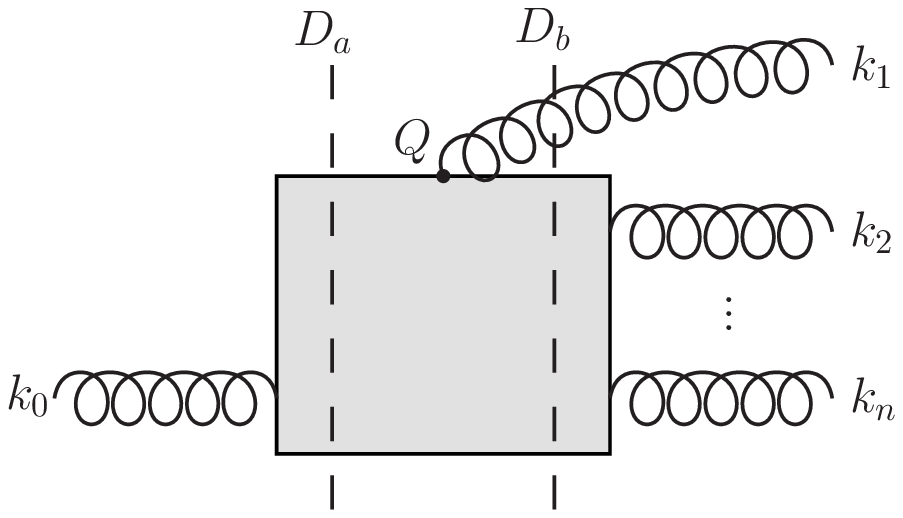}}}
\caption{Schematic representation of the crossing of the momentum of gluon 1 to relate the graphs for $1\to n+1$ with graphs $2\to n$. Point Q denotes the vertex at which gluon 1 attaches to the graph.  $D_a$ and $D_b$ denote the energy denominators for the intermediate states.}
\label{fig:denomgraphs}
\end{figure}

Let's consider the vertex factors $\cal V$.   By taking a graph and changing the direction of the $k_1$ line   only one vertex is changed, the one with the $k_1$ line ($v_1$).  The other vertices remain the same and, because the expressions for the internal lines do not change, neither do the expression for these vertices.  However, the expression for $v_1$ does change: it has an extra minus sign,  i.e., $v_{1,\: 2\to n}=-v_{1,\,1\to n+1}$. Therefore the vertex factor contains an extra sign,
${\cal V}_{2 \to n}=-{\cal V}_{1\to n+1}$.
Finally, let us consider   the energy denominators $\cal{D}$.  Let us consider a general $2\to 2$ graph with arbitrary number of vertices and internal lines, the graph can also contain loops, it is schematically depicted in Fig. \ref{fig:denomgraphs}.   Let $Q$ denote the point where gluon 1 interacts with the rest of a graph.  Using it as a reference, we split the graph into a left sub-graph and a right sub-graph.  Next, let denote by $D_a$ and $D_b$ the  energy denominators of the intermediate states $a$ and $b$ which are present just next to the vertex, we thus assume that there are more vertices to the left and right of the vertex $Q$, such that a and b are the intermediate states.  For Fig. \ref{fig:1ton+1denom} we can write
\be
D_a=E_a-\frac{\uvec{k}_{0}^2}{z_0 }=E_a+\frac{\uvec{k}_1^2}{z_1 }-\sum_{i=2}^n\frac{\uvec{k}_{i}^2}{z_i }=E_a-\frac{\uvec{k}_{A}^2}{z_A }-\sum_{i=2}^n\frac{\uvec{k}_{i}^2}{z_i } \; ,
\ee
and
\be
D_b=E_b-\frac{\uvec{k}_{0}^2}{z_0 }-\frac{\uvec{k}_{1}^2}{z_1 }=E_b-\frac{\uvec{k}_{0}^2}{z_0 }+\frac{\uvec{k}_{A}^2}{z_A }=E_b-\sum_{i=2}^n\frac{\uvec{k}_{i}^2}{z_i } \; ,
\ee
where $E_a$ and $E_b$ are the sums of the energies in intermediate states $a$ and $b$ respectively.  Here, we've used \eqref{eq:physconds} and explicitly excluded the energy of gluon 1 from $E_a$.  Similarly, for Fig. \ref{fig:2tondenom} we can write
\be
D_a=E_a-\frac{\uvec{k}_{0}^2}{z_0 }=E_a-\frac{\uvec{k}_1^2}{z_1 }-\sum_{i=2}^n\frac{\uvec{k}_{i}^2}{z_i }=E_a-\frac{\uvec{k}_{A}^2}{z_A }-\sum_{i=2}^n\frac{\uvec{k}_{i}^2}{z_i } \; ,
\ee
\be
D_b=E_b-\frac{\uvec{k}_{0}^2}{z_0 }+\frac{\uvec{k}_{1}^2}{z_1 }=E_b-\frac{\uvec{k}_{0}^2}{z_0 }+\frac{\uvec{k}_{A}^2}{z_A }=E_b-\sum_{i=2}^n\frac{\uvec{k}_{i}^2}{z_i } \; .
\ee
This time we've excluded the energy of the gluon 1 from $E_b$.  Looking at $D_a$ and $D_b$ for both cases  once again one sees that, using $\uvec{k}_A$ and $z_A$, the expressions are the same, though in each of these cases $\uvec{k}_A$ and $z_A$ are defined differently.  Hence, the contribution from energy denominators $\cal D$ also does not change.  This result is true for all graphs in light-front theory with arbitrary topology. We can therefore conclude that the $1\to n+1$ and $2\to n$ scattering amplitudes satisfy the relation 

\begin{align}
M_{2\to n}(\{\uvec{k}_0,z_0;\uvec{k}_1,z_1\};\{\uvec{k}_2,z_2;\ldots;\uvec{k}_n,z_n\})&=-M_{1\to n+1}(\{\uvec{k}_0,z_0\};\{\uvec{k}_A,z_A;\uvec{k}_2,z_2;\ldots;\uvec{k}_n,z_n\})\vert_{\uvec{k}_A\to-\uvec{k}_1,\:z_A\to-z_1}.
\label{eq:mainresult}
\end{align}
Furthermore, since the physical conditions for both cases are the same when written in terms of $\uvec{k}_A$ and $z_A$, we can use them to simplify $M_{2\to n}$ by simplifying $M_{1\to n+1}$ first as much as possible, which we are about to do.

The usefulness of \eqref{eq:mainresult} is that it converts a complicated problem into a simpler one (in addition, we get two results for the price of one).  Generally, $M_{1\to n+1}$ is  easier to calculate than $M_{2\to n}$, as we saw previously there are smaller number of graphs to compute.  Here we will see that it is a relatively simple task to show that $M_{2\to n}=0$ for the configuration of helicities $(-,+,+,\cdots,+)$ using \eqref{eq:mainresult} and the recursion relations derived earlier for the wave functions.  All one needs to do is to show that $M_{1\to n+1}=0$.  A substantial step towards this was done in \cite{ms}, and it has been summarized in previous section. Note that, unlike the graph shown in Fig.~\ref{fig:onium_n},  the external lines $1,2,\dots,n+1$ in $M_{1\to n+1}$ will be put on-shell, i.e. there is no energy denominator for the final state.  We thus obtain 
\be
M_{1 \to n+1} (k_0,k_1,k_2,\ldots, k_{n+1}) = 
{-2g}\, \sum_{i=2}^{n+1} \sqrt{z_0 z_1 z_2 \ldots z_{(i\;i+1)}\ldots z_{n+1}}
{\uvec{\epsilon}^{(+)} \uvec{v}_{i-1 \, i} }\, \Psi_{n} (k_1,\ldots ,k_{i-1 \, i},\ldots, k_{n+1})\; ,
\label{eq:recurrence1p}
\ee
where the square root eliminates the extra factors introduced by \eqref{eq:vnorm}.
There is also an overall summation performed over all the possible splittings of $n$ gluons of the original wave function.
$\Psi_{n}$ is given by equation (\ref{eq:psinfact})
\[
\Psi_{n}(1,2,\ldots,n) \; = \; (-1)^{n-1}(ig)^{n-1}\,\Delta^{(n)}\, {\sqrt{z_0} \over \sqrt{z_1 z_2 \ldots z_{n}}}\,
{1\over\xi_{(12\ldots n-1)n}\,\xi_{(12\ldots n-2)(n-1\,n)} \,
\ldots \, \xi_{1(2\ldots n)}}
\]
\be
\times\;
{1\over v_{(12\ldots n-1)n}\, v_{(12\ldots n-2)(n-1\,n)} \,
\ldots \, v_{1(2\ldots n-1)}} \; . 
\label{eq:psinfactn1}
\ee
Grouping the terms in (\ref{eq:recurrence1p}) to the common denominator using the explicit expression (\ref{eq:psinfactn1})  one arrives at the following formula \cite{ms} (which should be compared with \eqref{eq:fullsplit} and \eqref{eq:psinfact} )
\begin{align}
M_{1\to n}(k_0,k_1,\ldots,k_n)&= \\
 \; & (-1)^{n}i^{\;n-1} g^{n}\,{2\sqrt{z_0} \sum_{i=1}^{n} v^*_{i\,i+1}\xi_{(1\dots i)(i+1\dots n+1)} v_{(1\dots i)(i+1\dots n+1)}\over \,\xi_{(12\ldots n)n+1}\,\xi_{(12\ldots n-1)(n\,n+1 )} \,
\ldots \, \xi_{1(2\ldots n+1)}}  \;{1\over v_{(12\ldots n)n+1}\, v_{(12\ldots n-1)(n\,n+1 )} \,
\ldots \, v_{1(2\ldots n+1)}} \; .
\label{eq:sumnums}
\end{align}
Using \eqref{eq:sumvxi} one can show that the sum in the numerator coincides with the difference of the energies
\be
 \sum_{i=1}^{n} v^*_{i\,i+1}\xi_{(1\dots i)(i+1\dots n+1)} v_{(1\dots i)(i+1\dots n+1)}={1 \over 2}\left(\frac{\uvec{k}_{0}^2}{z_0} - \sum_{i=1}^n \frac{\uvec{k}_{i}^2}{z_i}\right) \; .
\label{eq:numerator}
\ee
The right-hand side of (\ref{eq:numerator}) is proportional to the difference of the  incoming  light-front energies and  the outgoing  light-front energies, which, according to the physical conditions of the problem, are the same.  Hence, the numerator in \eqref{eq:sumnums} is equal to zero, which gives $M_{1\to n+1} = 0 \to M_{2\to n}\sim P^{-}_{in}-P^{-}_{fin}=0$. Note that in arriving to the  result for the resummed wave function (\ref{eq:psinfactn1}) one utilizes the above relation \eqref{eq:numerator} for the last intermediate state. In such case the sum (\ref{eq:numerator}) is non-zero and it cancels with the denominator giving the non-vanishing resummed relation (\ref{eq:psinfactn1}).  Therefore we see how the recursion relations for the ($+; ++\dots+$) wave functions  are related to    the vanishing of the MHV amplitudes with helicities $(-,+,+,\dots,+)$ .

On the other hand the vanishing of the amplitudes $m(+,+,\dots,+)$ on the light-front can be argued from the angular momentum conservation as follows (see  \cite{Brodsky:2010ev}).
  
   The light-front wave functions satisfy non-trivial conservation of the total angular momentum
\be
J^z=\sum_{i=1}^n \, s_i^z + \sum_{j=1}^{n-1} \, l_k^z \;. 
\label{eq:jconservation}
\ee
The superscripts $z$ denote the projection along the light-front direction $z$. The first term corresponds
to the contribution  originating from the spin of the $n$ component Fock state of the wave function in question, whereas the second term is the sum over the relative orbital angular momenta. The second term excludes the contribution from the motion of the center of mass and is only due to the relative motion. The sum rule \eqref{eq:jconservation} has been verified for in QED for a systems of electron and spin 1(0) boson \cite{Brodsky:2000ii} and in QCD for dressed quark and gluon \cite{Harindranath:1998ve,Ma:1998ar}.
  For a transition of $2$ to $n$ the relative change of the contribution from
intrinsic spin is equal to $\Delta S_z=n+2$. On the other hand for tree-level  $g^n$ amplitude there are only
$n$ triple gluon vertices, each of which can change the angular momentum by $\Delta l_z \pm 1$. Thus the orbital angular momentum can only be changed at most by $\Delta L_z=n$. Thus the amplitude $m(+,+,\dots,+)$ needs to vanish on the light-front. Moreover the vanishing has to occur graph by graph in perturbation theory.

The same argument for the helicity flip amplitude $m(-,+,+,\dots,+)$ considered above is not sufficient as the requirement \eqref{eq:jconservation} can be in principle satisfied for each graphs. Thus the contributions from  individual graphs on the light-front  do not vanish as we saw in the 
the simple example earlier. It is only the sum of graphs which vanishes, and the proportionality of the single helicity flip amplitude  to the vanishing denominator on the light-front suggests the kinematical constraint due to the Lorentz invariance, similarly to the case of the tree amplitudes in gravity \cite{Grisaru:1975bx}.
\subsection{The $2\to 2$ amplitude $m(-,-,+,+)$ obtained from wave function $\Psi_3(+;-,+,+)$ by using the recursion relation}

In this section we are going to see how the previous discussion on  the relations between wave functions and the amplitudes works in a less trivial  case. We shall  namely consider  the case of  (+$\to$ -++) wave function and corresponding (++$\to$++) amplitude. We shall work with off-shell objects, leaving the comparison with the on-shell Parke-Taylor amplitudes to the very end, when the on-shell condition will be imposed. We shall  label the wave function of the graph in Fig.\ref{fig:onium_n} as $\Psi^{+\ldots+}_n$ for $(1,\ldots,n)=(+,\ldots,+)$ and $\Psi^{-+\ldots+}_n$ for $(1,2,\ldots,n)=(-,+,\ldots,+)$ .  With this notation, we can write the amplitude $\bar{M}_{1\to n}(+\to-++\ldots+)$ as
\begin{align}
\bar{M}_{1\to n}(+\to-++\ldots+) =&\;\Psi^{+\ldots+}_{n-2}(123,4,\ldots,n)\times \sqrt{z_0 z_4 \ldots z_n \over z_{123}}\left(ig^2+ig^2{(z_{123}+z_3)(z_2-z_1)\over(z_3-z_{123})^2}\right)\nonumber \\
&\;+ \Psi^{+\ldots+}_{n-1}(12,3,\ldots,n)\times \sqrt{\frac{z_0 z_3\ldots z_n}{z_{12}}} \left(2g\;z_2\; v_{(12)1}\right)\nonumber\\
&\,+\Psi^{-+\ldots+}_{n-1}(12,3,\ldots,n)\times \sqrt{\frac{z_0 z_3\ldots z_n}{z_{12}}} \left(2g\; z_1\;v_{2(12)}^*\right)\nonumber \\
&\;+\sum_{i=2}^{n-1} \Psi^{-+\ldots+}_{n-1}(1,2,\ldots,i\;i+1,\ldots,n) \times\sqrt{\frac{z_0 z_1\ldots z_n}{z_{i\;(i+1)} z_i z_{i+1}}}\left(2g\;z_{i\;(i+1)}\;v_{(i+1)\;i}^*\right) \; ,
\label{eq:1tonwneghel}
\end{align}
\begin{figure}[h]
\centering
\subfloat[]{\label{fig:hela}\includegraphics[width=.45\textwidth]{{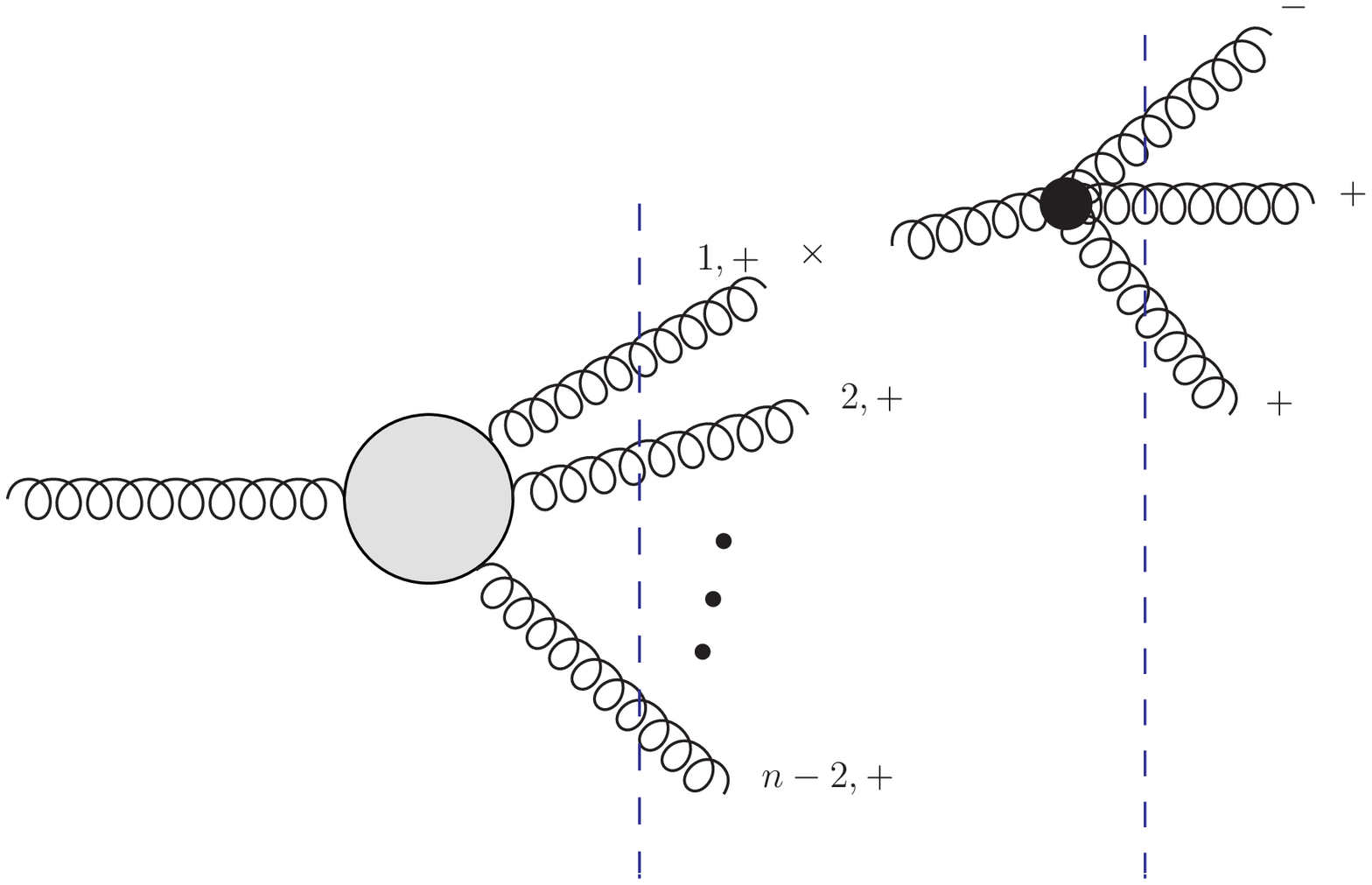}}}
\subfloat[]{\label{fig:helb}\includegraphics[width=.45\textwidth]{{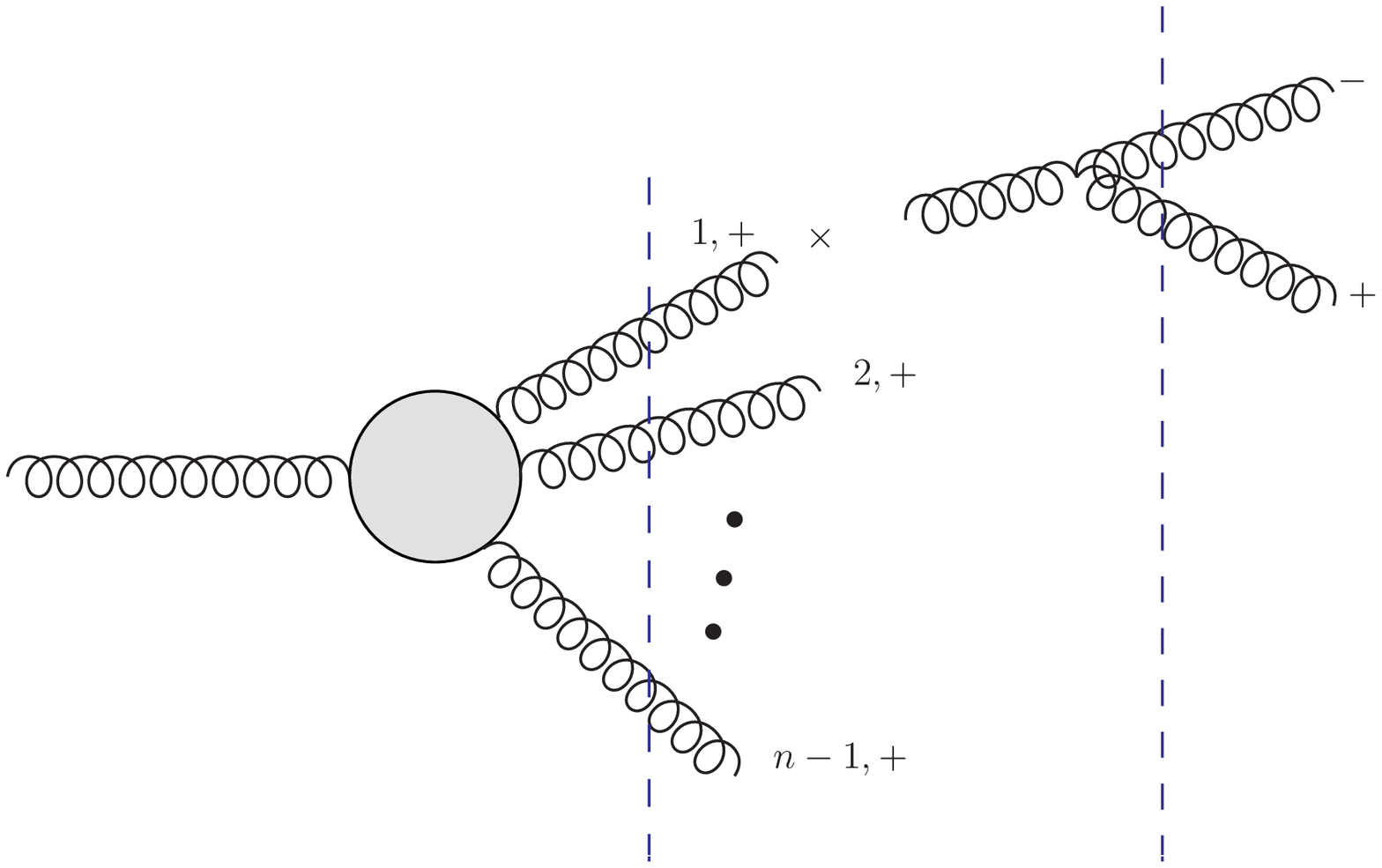}}}\\
\subfloat[]{\label{fig:helc}\includegraphics[width=.45\textwidth]{{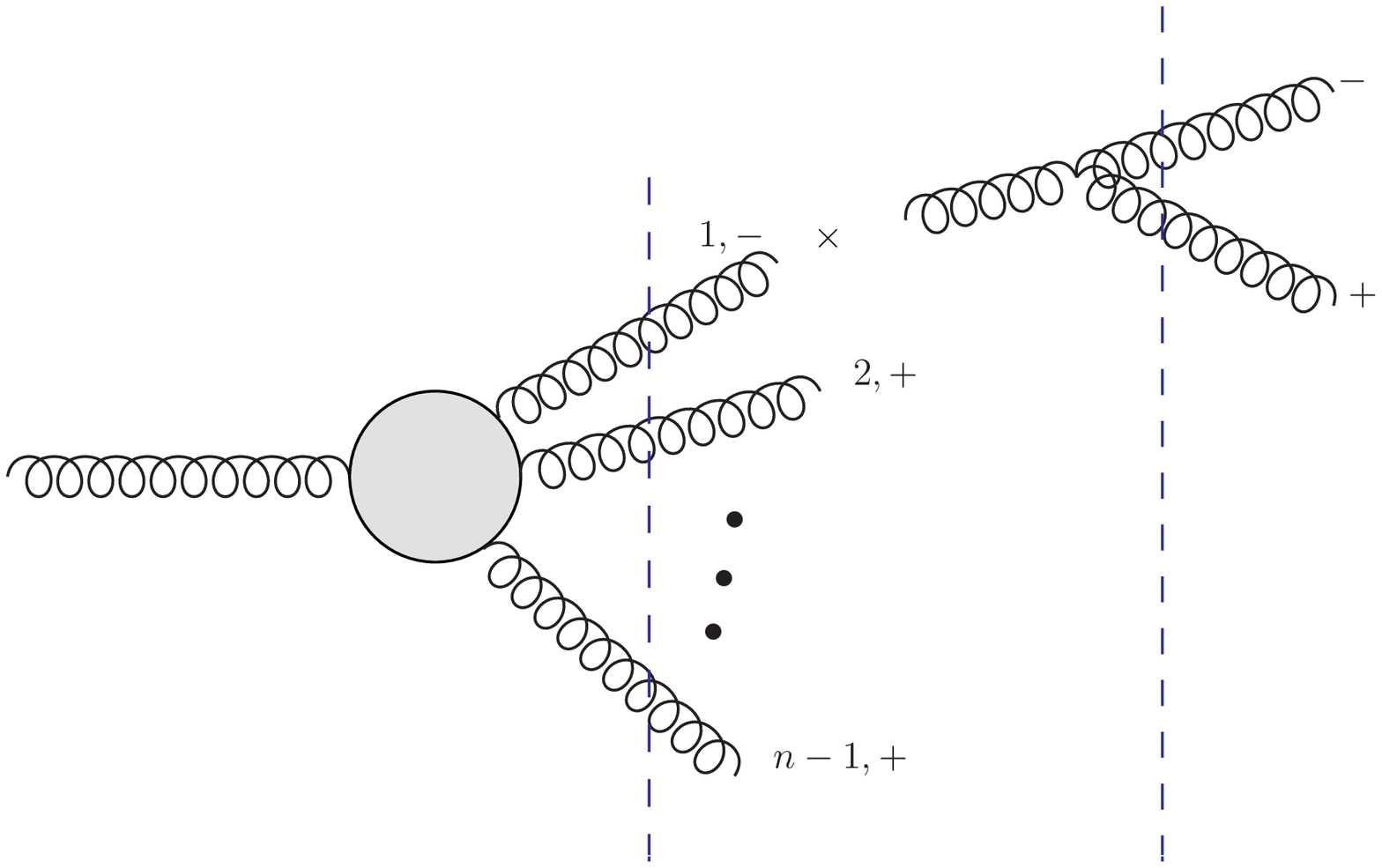}}}
\subfloat[]{\label{fig:held}\includegraphics[width=.45\textwidth]{{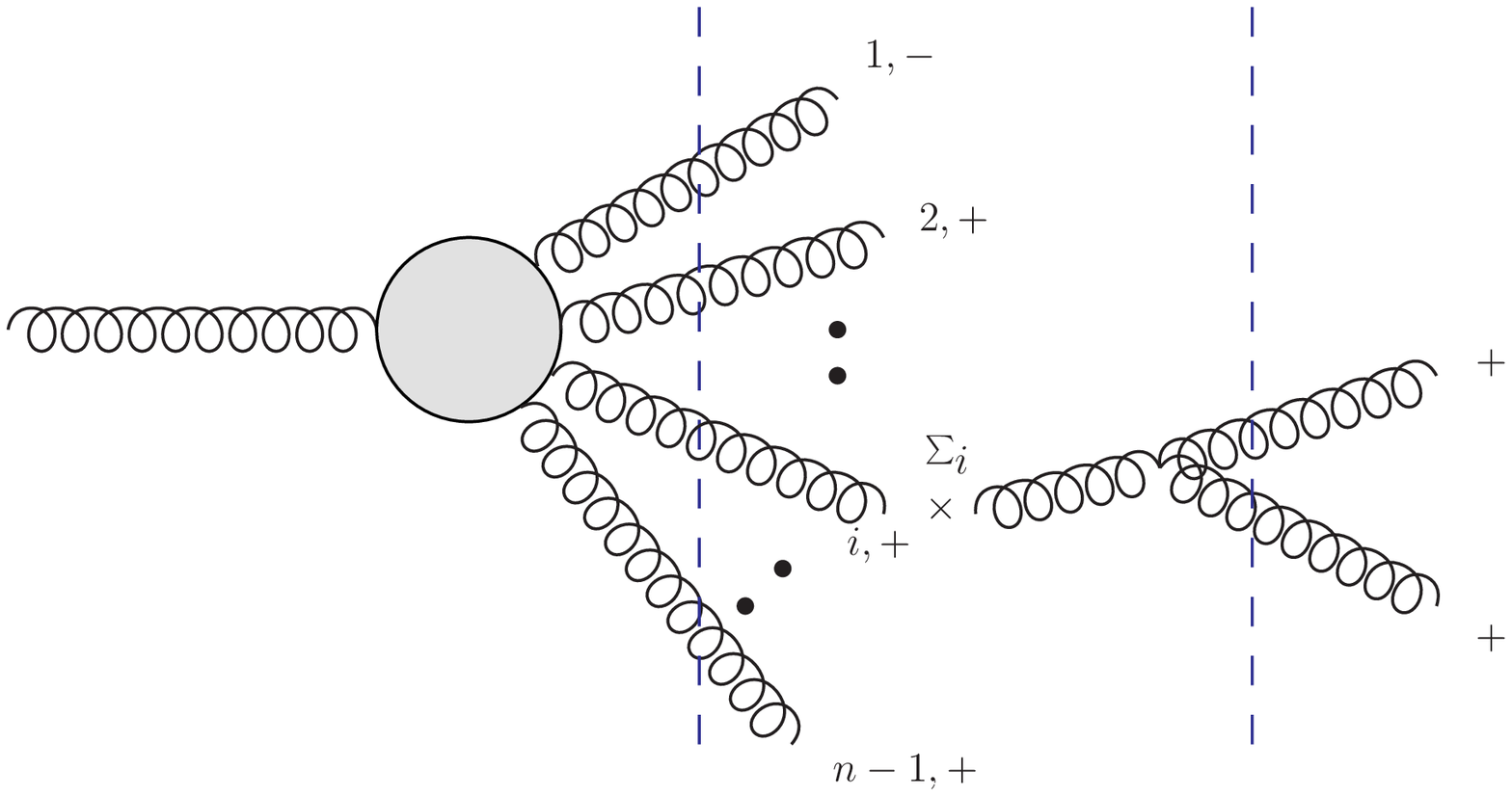}}}
\caption{A schematic representation of four terms on the right hand side of recursion \eqref{eq:1tonwneghel}. The blob in the four gluon vertex in graph (a) denotes the combined 4-gluon and Coulomb vertex. Vertical dashed lines indicate that the states are taken off-shell and the energy denominators are taken into account.}
\label{fig:recpsineghel}
\end{figure}
\begin{figure}[h]
\centering
\subfloat[]{\label{fig:4vertex}\includegraphics[width=.23\textwidth]{{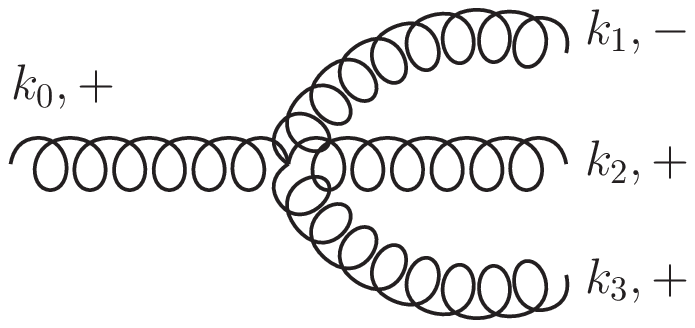}}}
\subfloat[]{\label{fig:instant}\includegraphics[width=.25\textwidth]{{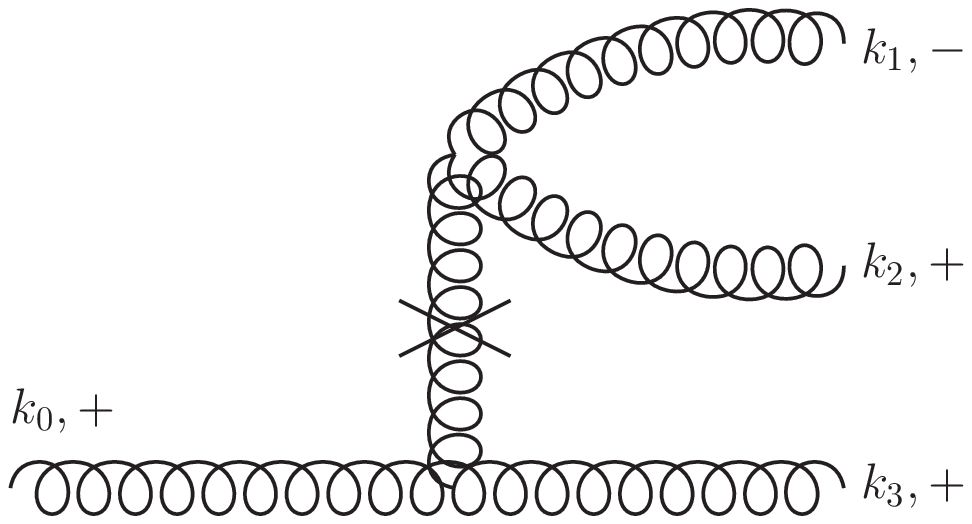}}}
\subfloat[]{\label{fig:pmp}\includegraphics[width=.23\textwidth]{{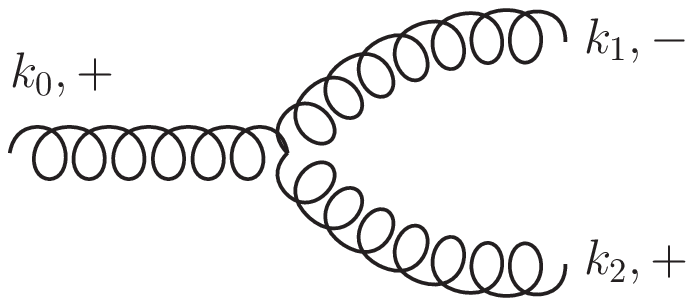}}}
\subfloat[]{\label{fig:mmp}\includegraphics[width=.23\textwidth]{{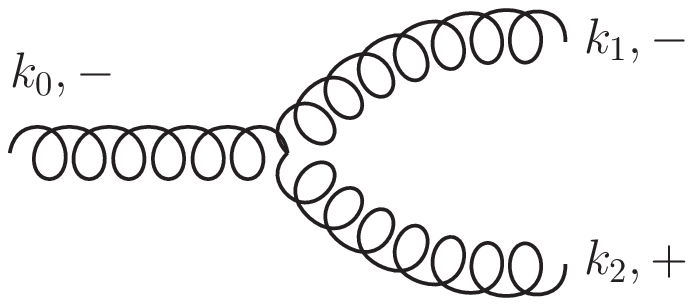}}}
\caption{Vertices that are included in recursion relation for the wave functions \eqref{eq:1tonwneghel}. The gluons on the left hand side of the vertices are treated as incoming.}
\label{fig:newvertices}
\end{figure}
where the   terms in parenthesis in the first three lines come from the vertices in Figs. \ref{fig:4vertex} - \ref{fig:mmp} respectively.  The recursion relation \eqref{eq:1tonwneghel}  is schematically shown in Fig.~\ref{fig:recpsineghel}. We introduce here notation $\bar{M}$ to denote that the on-shell condition on the last $n$ gluons has not been yet applied. The on-shell amplitudes will be denoted by $M$. In the case when   the energy condition is not applied, then we also can find the gluon wave function $\Psi^{-+\ldots+}_n$ from $\bar{M}_{1\to n}(+\to-++\ldots+)$ by the following relation
\be
\Psi^{-+\ldots+}_n(1,\ldots,n)={-i\over \sqrt{z_0 z_1 \ldots z_n} D_n} \bar{M}_{1\to n}(+\to-++\ldots+) \; .
\label{eq:psimnghel}
\ee
Eqs.~\eqref{eq:1tonwneghel} and \eqref{eq:psimnghel} form a recursion relation for the wave functions $\Psi^{-+\ldots+}_n$. The first term in Eq.~\eqref{eq:1tonwneghel} contains both the 4-gluon vertex and the Coulomb term (which are depicted in Fig.~\ref{fig:newvertices}), which have been combined to one effective vertex. The recursion relation  \eqref{eq:1tonwneghel} is much more complicated in form 
then the recursion relation for $\Psi^{++\ldots+}_n$ written previously, compare Eq.~\ref{eq:recurrence1}.  The complication stems from the fact that now the recursion involves two distinct wave functions for different set of helicities $\Psi^{-+\ldots+}_m$ and $\Psi^{++\ldots+}_m$ as well
as for different values of $m$ due to the presence of the 4-gluon vertex and the Coulomb term. 
This makes searching for the solution to this recursion for general $n$ much more involved than in the previous case when the recursion only involved the same type of objects i.e. $\Psi^{++\ldots+}_m$. 
 
For now let us therefore demonstrate that the general relation derived in the previous section works for the  (+$\to$ -++) case and that it coincides with the well known results obtained in the literature previously. For this case  we have
\begin{align}
M_{1\to3}(+\to-++)=&\;\Psi^+_1(123)\times \sqrt{z_0\over z_{123}} \frac{2ig^2}{(z_0-z_3)^2}(z_0 z_2-z_1 z_3)+\Psi^{++}_2(12,3)\times \sqrt{\frac{z_0 z_3}{z_{12}}} \left(2g\;z_2\; v_{(12)1}\right) \nonumber \\ 
&\;+\Psi^{-+}_2(12,3)\times \sqrt{\frac{z_0 z_3}{z_{12}}} \left(2g\; z_1\;v_{2(12)}^*\right)+\Psi^{-+}_2(1,23)\times\sqrt{\frac{z_0 z_1}{z_{23}}}\left(2g\;z_{23}\;v_{32}^*\right) \;.
\label{eq:mpp}
\end{align}
The $\Psi_1^+$ and $\Psi_2^{++}$ wave functions have already been computed and their expressions are given by Eq.~\eqref{eq:psinfact}. The new object $\Psi_2^{-+}$  can be found by multiplying the vertex in Fig. \ref{fig:pmp} by $-i\over \sqrt{z_0 z_1 z_2} D_2$:
\be
\Psi_2^{-+}(1,2)=ig \frac{1}{\sqrt{\xi_{(\bar{2}\bar{1})1}}}\frac{1}{\xi_{(\bar{2}\bar{1})1}v_{(21)1}^*}.
\ee
In addition, we can combine the following terms 
\be
\Psi^+_1(123)\times \frac{2ig^2}{(z_0-z_3)^2}z_0 z_2+\Psi^{++}_2(12,3)\times \sqrt{\frac{z_0 z_3}{z_{12}}} \left(2g\;z_2\; v_{(12)1}\right)=\Psi^{++}_2(12,3)\times \sqrt{\frac{z_0 z_3}{z_{12}}} \left(2g\;z_2\; v_{01}\right),
\ee
\be
\Psi^+_1(123)\times \frac{2ig^2}{(z_0-z_3)^2}(-z_1 z_3)+\Psi^{-+}_2(12,3)\times \sqrt{\frac{z_0 z_3}{z_{12}}} \left(2g\; z_1\;v_{2(12)}^*\right)=\Psi^{-+}_2(12,3)\times \sqrt{\frac{z_0 z_3}{z_{12}}} \left(2g\; z_1\;v_{23}^*\right).
\ee
Putting in the explicit expressions for  $\Psi$'s the following expression is obtained
\begin{align}
\bar{M}_{1\to3}(+\to-++)=&\;2ig^2\left[\frac{z_0 z_2}{z_3 z_{12}}\frac{v_{01}}{v_{30}}+\frac{z_1 z_3}{z_0 z_{12}}\frac{v_{23}^*}{v_{30}^*}+\frac{z_{23}^2}{z_0 z_1}\frac{v_{32}^*}{v_{01}^*}\right]\\
=&\;\frac{2ig^2}{z_0 z_1 z_3 z_{12}}\frac{1}{v_{30} v_{30}^* v_{01}^*}\left[z_1 z_2 z_0^2\; v_{01} v_{01}^* v_{30}^* + (z_1 z_3)^2 \;v_{23}^* v_{30} v_{01}^* + z_{23}^2 z_{12} z_3\;v_{32}^* v_{30} v_{30}^* \right],
\end{align}
where we used $z_{12} v_{3(12)}=z_0 v_{30}$ and $z_{12} v_{0(12)}^*=z_3 v_{30}^*$ to get the first line.  We can now do some manipulations to the third term of the second line:
\begin{align}
z_{23}^2 z_{12} z_3\;v_{32}^* v_{30} v_{30}^*=&\;z_{23}z_{12} z_3 z_2 v_{32}^*v_{32}^*v_{30}+z_{23}z_{12} z_3 z_1 v_{32}^*v_{01}^*v_{30} \\
=&\;z_1 z_2 z_3 z_{12} v_{32}^*v_{32}^*v_{01}+z_2^2 z_3 z_{12} v_{32}^*v_{32}^*v_{32}+(z_1 z_3)^2 v_{32}^*v_{01}^*v_{30}+z_0 z_1 z_2 z_3 v_{32}^*v_{01}^*v_{30}\\
=&\;z_1 z_2 z_3 z_{12} v_{32}^*v_{32}^*v_{01}+z_2^2 z_3 z_0 v_{32}^*v_{30}^*v_{32}+z_2^2 z_3 z_1 v_{32}^*v_{12}^*v_{32} \nonumber \\
&\;+(z_1 z_3)^2 v_{32}^*v_{01}^*v_{30}+z_0 z_1 z_2 z_3 v_{32}^*v_{01}^*v_{30}
\end{align}
where we have used the following relations $z_{23}v_{30}=z_1 v_{01}+z_2 v_{32}$, $z_{12}v_{32}^*=z_0 v_{30}^*+z_1 v_{12}^*$ and $z_{23} z_{12} = z_1 z_3 + z_0 z_2$ in order to rewrite the expressions.   Thus,
\begin{align}
\bar{M}_{1\to3}(+\to-++)=&\;\frac{2ig^2}{z_0 z_1 z_3 z_{12}}\frac{1}{v_{30} v_{30}^* v_{01}^*}[z_1 z_2 z_3 z_{12} v_{32}^*v_{32}^*v_{01}+\left(z_1 z_2 z_0^2\; v_{01} v_{01}^* v_{30}^* +z_2^2 z_3 z_0 v_{32}^*v_{30}^*v_{32}\right) \nonumber \\
&\;+\left(z_2^2 z_3 z_1 v_{32}^*v_{12}^*v_{32}+z_0 z_1 z_2 z_3 v_{32}^*v_{01}^*v_{30}\right) ]
\end{align}
Finally, we can group some terms which are proportional to denominator $D_3$
\be
z_1 z_2 z_0^2\; v_{01} v_{01}^* v_{30}^* +z_2^2 z_3 z_0 v_{32}^*v_{30}^*v_{32}=-\frac{1}{2}z_0 z_2 z_{23} v_{30}^*D_3,
\ee
\be
z_2^2 z_3 z_1 v_{32}^*v_{12}^*v_{32}+z_0 z_1 z_2 z_3 v_{32}^*v_{01}^*v_{30}=-\frac{1}{2} z_1 z_2 z_3 v_{32}^* D_3.
\ee
Therefore,
\be
\bar{M}_{1\to3}(+\to-++)=\frac{2ig^2}{z_0 z_1 z_3 z_{12}}\frac{1}{v_{30} v_{30}^* v_{01}^*}\left[z_1 z_2 z_3 z_{12} v_{32}^*v_{32}^*v_{01}-{1\over2}z_2D_3\left(z_1 z_3 v_{32}^*+z_0 z_{23}v_{30}^*\right)\right]
\label{eq:barmfinal}
\ee
The on-shell energy condition $D_3=0$ can now be applied to the above expression as required for the on-shell amplitudes. This leaves only the first term in Eq.~\eqref{eq:barmfinal} which is finite and it  matches the Parke-Taylor amplitude. This can be made more explicit by utilizing the relation
$z_0 v_{01}^* v_{30}=-z_2 v_{12}^* v_{32}$ and $z_0z_1v_{01}^*v_{01}=z_2 z_3 v_{23}^* v_{23}$ which leads us to the expression

\be
M_{1\to3}(+\to-++)=\frac{2ig^2 (z_0 z_1)^2}{z_0 z_1 z_2 z_3}\frac{v_{01}^4}{v_{01} v_{12} v_{23} 
v_{30} }
\ee
 which is the exact result for this case.
 Thus by utilizing the recursion relation for the wave functions we are able to reconstruct the 
 the on-shell Parke-Taylor amplitude. The full proof would of course involve solving exactly the recursion relation, which we leave for a future work. An interesting structure though emerges from this example, namely the fact that the off-shell amplitude $\bar{M}$ can be written as a sum of the 
 on-shell amplitude plus the term proportional to the energy denominator which vanishes 
 if one takes the on-shell condition $D_3\rightarrow 0$. Thus we can expect that the general structure for arbitrary number of final particles in the LFPT has the following form
 \be
 \bar{M}(2\rightarrow n) ={M}(2\rightarrow n) + {\cal{O}}(D_n) \;,
 \ee
 where the term which is related to the off-shellness of the amplitude vanishes upon taking the physical
 condition $P^-_{in}=P^-_{fin}$, i.e. it is at least linear in the energy denominator $D_n$. 
\section{Conclusions and outlook}

In this paper we have investigated the gluon wave functions, fragmentation functions and scattering amplitudes within the framework of the light front perturbation theory.  The recursion relations for each of these objects have been constructed on the light-front. For the special case, when the helicities of the outgoing gluons are the same, it is possible to solve the recursion relations for the gluon wave functions and fragmentation functions and obtain the final results  in a very compact form. Furthermore, we have shown  that in general the scattering amplitudes (with off-shell final states) can be computed from the gluon wave functions. One of the advantages of using the LFPT is the fact that the variables used to express the helicity amplitudes naturally arise in this framework.

In particular, we have shown that the compact recursion relations for the gluon wave functions derived before for the special case of the helicity configuration  are  related to the vanishing property of the on-shell
helicity amplitudes for selected configurations of the helicities.  We have generalized  the recursion relations for the wave functions and the fragmentation functions to include general configurations of helicities. In the latter case, the recursion relations are light-front analogs of the previously used Berends-Giele recursion relations. We have also verified the vanishing of the amplitudes on the light-front for special cases of the helicities, which originates from the angular momentum conservation and energy conservation. Using the  general relations between and gluon wave functions and  scattering amplitudes we have been also able to reproduce some of  the results available in the literature. The light-front methods presented here can be used to compute the scattering amplitudes with arbitrary number of external legs and for different helicity configurations.

\section*{Acknowledgments}
We would like to thank Stan Brodsky for discussions.
 This work was supported  in part  by the DOE OJI grant No. DE - SC0002145   and by  the   Polish NCN 
grant DEC-2011/01/B/ST2/03915.  A.M.S. is supported by the Sloan Foundation.

\section*{Appendix}

\begin{figure}[h]
\centering
\subfloat[]{\label{fig:appA}\includegraphics[width=.5\textwidth]{{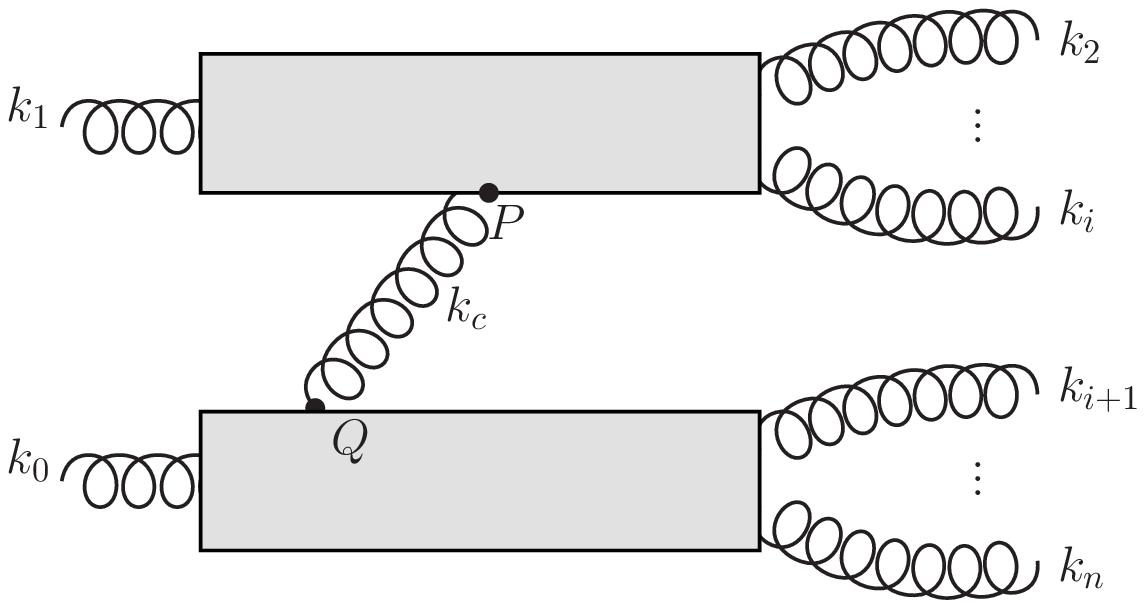}}}
\subfloat[]{\label{fig:appB}\includegraphics[width=.5\textwidth]{{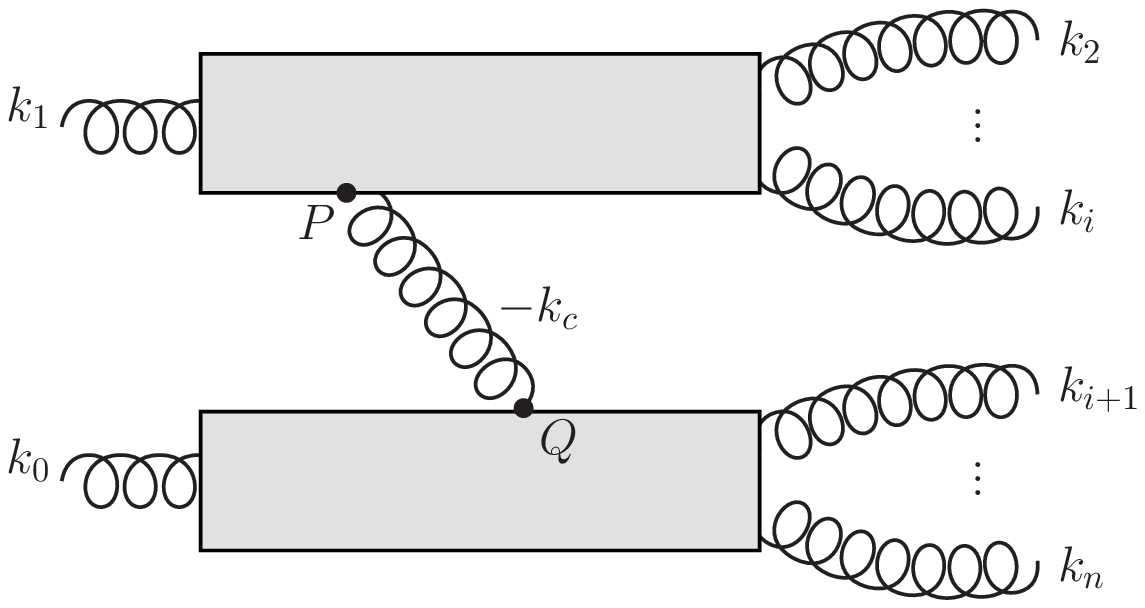}}}
\caption{Generic 2 to $n$ tree-level graphs. In (a) gluon $c$ departs from the lower subgraph at Q and arrives at the upper subgraph at P.  For (b) the opposite is true.}
\label{fig:app}
\end{figure}

In this Appendix we shall investigate a general structure of the tree-level amplitudes in light-front theory. In particular we shall demonstrate that the typical tree-level graphs can be written in terms of the factorizable contributions originating from initial and final state emissions and non-factorizable contribution which typically comes from the energy denominator of the intermediate state exchanged between two subsets of graphs. On top of that we shall show  
for a given group of topologically identical graphs, all the subgroups which differ by the orientation of the exchanged particle within such group have the same value. This is true with with the exception of those subgraphs for which the physical conditions constrain the graphs to have a value of zero (i.e. vacuum graphs). We will show this by demonstrating that graphs within the given subgroup sum up to the expression which can be analytically continued to obtain the graphs from the different subgroup.  Generic tree-level graphs are shown in   Fig.~\ref{fig:appA} and Fig.~\ref{fig:appB}.   In these figures, the top and bottom box both have a set of vertices and thus form subgraphs with a definite topology.  
The graph represents the sum of all the possible time-orderings between the upper and lower parts of the diagrams.

Points P and Q denote the vertices at which gluon labeled by $c$ attaches to both subgraphs. We need also to perform the sum over the relative orderings of these vertices with respect to the orderings in the lower and upper subgraphs. For instance, for situation depicted in Fig.\ref{fig:appA} we need 
to perform the sum over all the orderings of the splittings in the upper box  which occur before the vertex P with respect to the position of vertex Q. One can call them initial state splittings of the subgraph U, as they occur before the interaction P. Similarly, we have to sum over all the orderings of the splittings of lower subgraph L with respect to P which occur after interaction 
depicted by the vertex Q. These splittings one can refer to as the final state splittings of the subgraph L. A similar situation is valid for a graph shown in Fig.~\ref{fig:appB}.

\begin{figure}[ht]
\centerline{\includegraphics[width=0.6\textwidth]{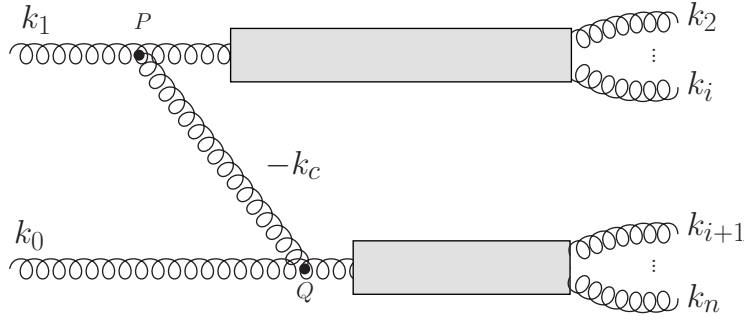}}
\caption{2 to $n$ tree-level graph with only final state splittings.}
\label{fig:appC}
\end{figure}

Let us consider a graph with only final state splittings, that is with the splittings which occur in the upper part occur only after  vertex  P and splittings in the lower subgraph occur only after vertex Q. Such graph is depicted in Fig.~\ref{fig:appC}.  Let us label by  $A_i$ the energy denominators of the upper subgraph U in the absence of  lower subgraph and gluon $c$, and by $B_j$ the energy denominators in the lower subgraph in the absence of the upper part as well as the gluon $c$. In particular $A_i(B_j)$ is the energy denominator of the intermediate state which is on the $i(j)$ position when we count from the right hand side of the corresponding subgraph. We denote by $C$ the energy of the intermediate gluon $c$. Consider a class of subset of graphs in which all the splittings in L and U occur after vertex P. It was shown in \cite{ms} that in such case the fragmentation trees factorize. Using this property for such configuration we can write the contribution from the energy denominators in the form
\be
\frac{1}{A_m-C+\bar{B}}\frac{1}{A_1A_2\dots A_{m}B_1\dots B_{n}} \; ,
\label{eq:fact1}
\ee
where by $\bar{B}$ we denoted the difference of the energy denominator of the gluon 0,i.e.
$\bar{B}=\bar{B}_0-B_0$ where $\bar{B}_0$ is the energy of the incoming gluon into the lower subgraph  L and $B_0$ is the sum of the  energies of the final states in the same subgraph L.

The next step is to include the remaining graphs of the same topology, but this time  with the splittings in the lower subgraphs occurring earlier than the point P but later than Q. Summing this subset of diagrams together with the diagrams which result in (\ref{eq:fact1}) one arrives at the factorized expression. Again, the upper tree factorizes from the lower part and the sum of all the time-orderings for the graphs in Fig.~\ref{fig:appC} is 
$$
\frac{1}{-C+\bar{B}}\frac{1}{A_1A_2\dots A_{m}B_1\dots B_{n}} \; ,
$$

\begin{figure}[ht]
\centerline{\includegraphics[width=0.6\textwidth]{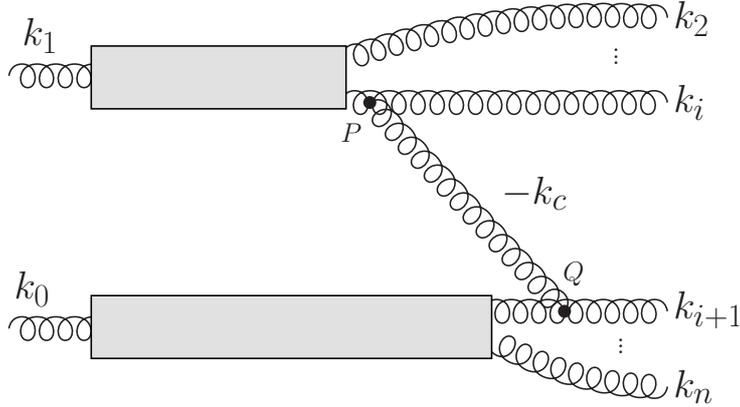}}
\caption{2 to $n$ tree-level graph with only initial state splittings.}
\label{fig:appD}
\end{figure}

The first factor $\frac{1}{-C+\bar{B}}$ is just the energy denominator for the simple $2 \to 2$ graph, without any splittings in the final state, with the only assignment that the final energy of the lower leg in this graph is given  by $B_0$. This denominator in turn  multiplies  the factorized contributions to the fragmentation trees originating from subgraphs L and U. The  same procedure can be repeated for the general graph depicted in Fig.~\ref{fig:appD} with only the  initial state radiation. Finally one needs to consider the most general case in which the splittings can occur both before and after P and Q on both the upper and lower subgraphs simultaneously.  In this case the contribution from the energy denominators is given by
\be
\frac{1}{\bar{A_1}\bar{A}_2\dots\bar{A}_k\bar{B_1}\bar{B}_2\dots\bar{B}_l}\frac{1}{-C+\bar{B}}\frac{1}{A_1A_2\dots A_{m}B_1\dots B_{n}} \; .
\label{eq:fact_tot}
\ee
A similar result can be obtained for the graph Fig.~\ref{fig:appA} for which the contribution from the  energy denominators reads
$$
\frac{1}{\bar{A_1}\bar{A}_2\dots\bar{A}_k\bar{B_1}\bar{B}_2\dots\bar{B}_l}\frac{1}{-\tilde{C}+\bar{A}}\frac{1}{A_1A_2\dots A_{m}B_1\dots B_{n}} \; .
$$
where now $\bar{A}$ is the difference between the energy of the incoming gluon 1 in the upper subgraph U and the sum of the final energies of the states in the same subgraph U.  Using global light-front energy conservation $\bar{A}=-\bar{B}$ and performing substitution $C=-\tilde{C}$ we see that the contribution from the energy denominators for both graphs is the same up to a global sign. One needs to take into account also the additional sign change between graphs \ref{fig:appB} and \ref{fig:appA} coming from the contributions  from the longitudinal fractions. Thus the general expression for the graph (either \ref{fig:appA} or \ref{fig:appB}) has a compact (partially) factorizable form with the energy denominators given by expression (\ref{eq:fact_tot}).


\bibliographystyle{elsarticle-num}
\bibliography{<your-bib-database>}



\end{document}